\documentclass[preprint,amsmath,amssymb, nofootinbib]{revtex4}
\usepackage{graphicx}
\usepackage[dvipsnames]{xcolor}
\usepackage{subcaption}
\usepackage{tabularx}
\usepackage{multirow}
\usepackage{array}
\usepackage{makecell}
\usepackage{dcolumn}
\usepackage{array}
\usepackage{footnote}
\makesavenoteenv{table} 

\usepackage{adjustbox}
\usepackage{hyperref}
\usepackage{bm}

\begin{document}

\title{Photon rings in the metamaterial analog of a gravitomagnetic monopole}
\author{${\rm A.\;Parvizi}^{\;(a,b,c)}$ \footnote{Electronic
address:~aliasghar.parvizi@uwr.edu.pl}, ${\rm H. \;\rm Forghani}$-${\rm Ramandy}^{\;(a)}$\footnote{Electronic
address:~hasan.forghani@ut.ac.ir}, ${\rm E.\;Rahmani}^{\;(a)}$\footnote{Electronic
address:~rahmani.elnaz@ut.ac.ir} and ${\rm M. \;Nouri}$-${\rm Zonoz}^{\;(a)}$ \footnote {Electronic
address:~nouri@ut.ac.ir\; (Corresponding author)}}
\affiliation{(a):Department of Physics, University of Tehran, North Karegar Ave., Tehran 14395-547, Iran.\\
(b):Institute for Research in Fundamental Sciences (IPM), Farmanieh, Tehran, Iran.\\
(c): Institute for Theoretical Physics, Faculty of Physics and Astronomy,
University of Wroclaw, pl. M. Borna 9, 50-204 Wroclaw, Poland.}

\begin{abstract}
After studying null geodesics in the equatorial plane in NUT spacetime, we show that there are unstable photon rings in this plane. Next we transform the metric of this plane to the isotropic coordinates and introduce its equivalent two-parameter index of refraction. Utilizing the analog gravity concepts, we assign this index of refraction to a metamaterial analog of this plane, and by ray-tracing simulation find its photon rings. Furthermore, we extend our analysis to the charged NUT spacetime and employ wave optics to numerically solve Maxwell’s equations for electromagnetic waves within an inhomogeneous medium assigned with this spacetime's three-parameter index of refraction. We investigate the optical properties of such  metamaterial analogs for designing enhanced and fine-tuned  optical devices.

\end{abstract}
\maketitle
\section{Introduction}
Metamaterial analogs of different spacetimes, based on their optical characteristics, have already been studied in the literature, specialy after the introduction of the transformation optics \cite{Pen, Leon}.
One of the main features of metamaterials, which was utilized through transformation optics, was the great ability in controlling their optical properties by assigning them with the desired optical parameters . Linking this with the  behaviour of light rays in the exact solutions of Einstein field equations, through the assigned index of refraction, one can in principle design metamaterial analogs of these spacetimes mimicking their interesting  optical  behavior.  These include the formation of photon spheres in the ray tracing simulations of metamaterial analogs of static spherically symmetric spacetimes \cite{NPF}, and wave optics simulations of light propagation in the metamaterial analogs of different spacetimes \cite{Green, Chen, Fer, Maslov, Ting}. Based on the same analogy people have investigated properties of the so-called optical black holes in their dielectric and metamaterial analogs \cite{Leon2, Nariman, Genov}.\\
This opto-geometric relation  provides us with the possibility of testing the optical properties of the most exotic solutions of Einstein field equations, namely black holes, in laboratory settings. On the other hand one can design metamaterial devices with interesting and unique optical features that utilize optical features in the corresponding black hole spacetime, for instance optical concentrator proposed in \cite{azevedo2018optical} or Hawking radiation in an optical analog of the black-hole horizon \cite{rosenberg2020optical}.\\
One of the interesting exact solutions of the Einstein field equations which has not been studied in this context is the NUT (Newman-Unti-Tamburino) solution \cite{NUT}. This solution, interpreted as the gravitational analog of a magnetic monopole, or the so called gravitomagnetic monopole, has a number of exotic properties \cite{Misner1}. These properties, originated from its Dirac-type string singularity, featured NUT spacetime as ``{\it a counter example to almost anything}'' \cite{Misner2}.
The NUT parameter plays a significant role in gravitational systems and the exact solutions of Einstein field equations (EFEs). It introduces a form of ``twist'' to the spacetime, which is not present in other well known solutions of the EEFs like the Schwarzschild or Kerr metrics. The presence of the NUT parameter leads to interesting features, such as the presence of the closed timelike curves, which challenge our conventional understanding of causality.
Interest in this spacetime in recent years has led to more studies including the study of its thermodynamics in four and higher dimensions \cite{Awad}-\cite{Clark}. It also played an important role in string theory when it was first embedded as a nontrivial stringy solution in the so called heterotic string theory \cite{John}. As in the case of other spacetimes, developing a metamaterial analog of the NUT spacetime could facilitate a better understanding of its properties through laboratory-scale experiments.\\
NUT spacetime is a stationary spacetime, and in the case of stationary spacetimes, unlike the static spherically symmetric spacetimes,  one can not
transform the corresponding metric to an isotropic coordinates to be able to assign an index of refraction to the whole spacetime. On the other hand in the case of NUT  solution, one could restrict the study to the equatorial plane $(\theta = \pi/2$) which could be transformed to the isotropic coordinates, and be assigned with an index of refraction. We will show that the main optical feature in this plane, i.e the presence of {\it photon rings} as unstable null orbits, could be implemented in its metamaterial analog by assigning the metamaterial with the plane's index of refraction. The closest case to the NUT spacetime which is both stationary and has two parameters, as well as possessing equatorial photon rings, is the Kerr spacetime \cite{Chandra}, so we will compare our results with those obtained in the Kerr case and its metamaterial analog \cite{Ting}.
Unlike the study in \cite{Ting}, we will find an exact simulations of rays in the metamaterial analog of NUT spacetime, and to a good approximation observe the same photon rings in the wave optics simulation of the (charged)  NUT spacetime. Another technical advantage of simulating photon rings in the NUT case, due to its two dimensional nature, is the much easier handling of the rays passing through the ring which we will discuss later. \\
The outline of the paper is as follows. In the next section we give a brief review of the NUT spacetime  and its interesting features. In section III we will discuss the presence of photon rings in  equatorial NUT spacetime, and transform its metric to isotropic coordinates, thereby assigning it with an index of refraction. In section IV we will employ the ray-tracing simulation in the  metamaterial analogs of the equatorial NUT and pure NUT spacetimes, and find their photon In Section V, we consider an extension of the NUT spacetime, the so-called {\it charged} NUT spacetime, which contains electric charge as its third parameter. Calculating its equivalent index of refraction, we employ a wave optics approach to explore the optical features of its metamaterial analog. We show that the inclusion of three parameters provides enhanced control over the radii of photon rings in the corresponding metamaterial analog, and consequently facilitate the design of optical devices based on them.\\
Conventions: In what follows we will use natural units in which $c=G =1$, and the metric signature is given by (1,-1,-1,-1).
\section{A Brief review of NUT spacetime}
NUT spacetime in Schwarzschild-like coordinates is given by the following metric \cite{Exact},
\begin{equation}\label{n1}
ds^2 = f(r) (dt - 2l \cos\theta d\phi)^2 - \frac{dr^2}{f(r)} - (r^2 + l^2) d\Omega^2
\end{equation}
with
\begin{equation}\label{f}
f(r) = \frac{r^2 - 2mr - l^2}{r^2 + l^2}
\end{equation}
in which $m$ and  $l$ are the mass and  NUT (magnetic mass)  parameters respectively. For $l=0$ we recover the Schwarzschild metric, whereas for $m=0$ unlike the Kerr case, we do not find flat spacetime in an exotic coordinate, but the so called {\it pure} NUT spacetime which is a one-parameter stationary spacetime. \\
Although, the NUT spacetime is mathematically axially symmetric with its axis along the {\it string singularity} at $\theta = 0, \pi$, its curvature invariants are spherically symmetric. This is due to the fact that one could transform the above metric, using the time coordinates
\begin{gather}
 t_N = t-2l\phi \label{tn} \\
 t_S = t+2l\phi \label{ts}
\end{gather}
to the following two forms respectively
\begin{gather}\label{n2}
ds^2 = f(r) \left(d{t_N} + 4l \sin^2 \frac{\theta}{2} d\phi\right)^2 - \frac{dr^2}{f(r)} - (r^2 + l^2) d\Omega^2  \\
ds^2 = f(r) \left(d{t_S} - 4l \cos^2 \frac{\theta}{2} d\phi\right)^2 - \frac{dr^2}{f(r)} - (r^2 + l^2) d\Omega^2 \label{n3},
\end{gather}
where in the first form there is a singularity only in the half-axis $\theta = \pi$, whereas in the second form the singularity is in the half-axis $\theta = 0$. This in principle shows that one could change the direction of the half-axis singularity to any other direction. Using this freedom Misner showed that one could remove the string singularity by covering the whole manifold with two different coordinate patches for the northern ($0 < \theta < \pi/2 $), and the southern ($\pi/2 < \theta < \pi $) regions. For $r=\rm constant$ hypersurfaces, these are solid tori, with topology $S^1 \times {E}^2$, sharing a boundary at $\theta = \pi/2$ which is a torus ($S^1 \times S_t$) \cite{Misner1}. To achieve this it is noted from \eqref{tn}-\eqref{ts}, that the identification of time at the boundary leads to
\begin{equation}
t_{N} = t_{S} - 4l\phi
\end{equation}
showing the angular character of either of the time coordinates. Consequently, by the single-valuedness of the $\phi$ coordinate, it leads to
\begin{equation}
t_{N/S} = t_{N/S} \pm 8\pi l.
\end{equation}
In other words the singularity-free form of the metric comes at the price of introducing a periodic time, but this shows that one could in principle find a singularity-free form of the metric which justifies the {\it physical} spherical symmetry of the NUT hole, despite its mathematically evident axially  symmetric metric.\\
Taking the radial coordinate $0 < r < \infty $, there is a coordinate singularity in NUT spacetime at $r_H = m + (m^2 + l^2)^{1/2}$ (where $f(r)=0$) as the NUT horizon which unlike the Schwarzschild horizon, does not hide a singularity at $r=0$, so we may call the source of the NUT spacetime either a NUT {\it hole} or a NUT {\it black hole}.\\
It is noted that as in the case of Schwarzschild spacetime inside the horizon, i.e for $0 < r < r_H $ where  $f(r) < 0$,
$r$ and $t$ change their role. This region is called the Taub region, and its metric which is non-stationary is the so called Taub metric (solution) discovered before the NUT spacetime \cite{Taub} as a spatially homogenous vacuum cosmological solution of Einstein field equations. This is why the metric including both the non-stationary and stationary regions, is called the Taub-NUT spacetime. In its maximal extension where $-\infty < r < \infty $, there are coordinate singularities at $r_{\pm} = m  \pm (m^2 + l^2)^{1/2}$ corresponding to two distinct NUT regions  for $r > r_+$ and $r < r_- $ with $f(r) > 0$,
separated by the Taub region $r_- <r < r_+$ where $f(r) < 0$ \cite{Miller}.
\\
Because of its Dirac-type singularity, the NUT space is interpreted as the spacetime of a mass endowed with a gravitomagnetic monopole charge (the NUT parameter). This justifies its spherical symmetry despite its axisymmetric appearance \cite{DN,DLBMNZ}. Bonnor gave another interpretation of NUT spacetime in terms of a semi-infinite massless source of angular momentum \cite{Bonnor}.
In what follows we will focus on the stationary region of the Taub-NUT spacetime, namely the NUT region ($r_H < r < \infty$), and its metamaterial analog.
\section{Photon rings in the equatorial NUT spacetime in isotropic coordinates}\label{sectionIII}
The null geodesics of NUT spacetime have already been studied in the literature, with an emphasis on light bending \cite{Zimm}-\cite{MNDL}. Here  we are interested in its photon rings in the equatorial plane, and in {\it isotropic} coordinates. To this end we restrict our attention to the line element of the subspace $\theta = \pi/2$, namely
\begin{equation}\label{n4}
ds^2 = f(r) dt^2 - \frac{dr^2}{f(r)} - (r^2 + l^2) d\phi^2,
\end{equation}
in which the metric function $f(r)$ is defined in \eqref{f}. The Lagrangian for the geodesics of the NUT spacetime is
\begin{equation}\label{L1}
2{\cal L }= f(r) \dot{t}^2  - \frac{\dot{r}^2}{f(r)} - (r^2 + l^2) {\dot{\phi}^2}
\end{equation}
where $^{.} \equiv d/d\lambda$ and $\lambda$ is an affine parameter along the null geodesics. Since the Lagrangian is independent of $t$ and $\phi$, we have the following two first integrals from the Euler-Lagrange equations,
\begin{equation}\label{L2}
f(r)\dot{t} = E
\end{equation}
\begin{equation}\label{L3}
(r^2 +l^2) \dot{\phi} = L
\end{equation}
representing the energy and angular momentum respectively. Also for null geodesics we have  $ds^2= 0 = {\cal L }$ leading to
\begin{equation}\label{L4}
f(r) \dot{t}^2  - \frac{\dot{r}^2}{f(r)} - (r^2 + l^2) {\dot{\phi}^2}=0
\end{equation}
Unstable circular null geodesics forming at radial coordinate $r=r_c$, are given by $\dot{r}=0$ in the above equation, which along with equations \eqref{L2}-\eqref{L3} leads to
\begin{equation}\label{L5}
\frac{\dot{t}}{\dot{\phi}} = \frac{L}{E} \equiv b_c
\end{equation}
in which $b_c$ by definition is the impact parameter for rays coming from infinity to be trapped on the photon ring at $r=r_c$. Replacing this back into equation \eqref{L4} for a photon ring we end up with
\begin{equation}\label{L6}
({r_c}^2 + l^2)^2 = ({r_c}^2 -2mr_c - l^2){b_c}^2,
\end{equation}
other equation governing the unstable circular photon rings can be derived by taking the derivative of the above equation with respect to $r_c$, which yields 
\begin{equation}\label{photonring2}
4 r_c \, ({r_c}^2 + l^2) = (2 {r_c} -2m){b_c}^2,
\end{equation}
then we read impact parameter $b_c$ as follows
\begin{equation}\label{L7}
{b_c} =\sqrt{\frac{2r_c({r_c}^2 + l^2)}{r_c - m}}
\end{equation}
Replacing this back into equation \eqref{L6}, we find the following 3-rd order equation for the radial coordinate of the photon ring
\begin{equation}\label{L8}
{r_c}^3 - 3 m {r_c}^2 - 3 l^2 r_c + m l^2 = 0
\end{equation}
with the negative discriminant $\Delta = - l^2 \left(l^2+m^2\right)^2$, all the three roots are real and unequal,
\begin{eqnarray}\label{eq:solutions}
    r_c &=& m+2 \sqrt{l^2+m^2} \cos \left(\frac{1}{3} \tan^{-1} (\frac{l}{m})\right)\label{L9}, \label{eq:solutions1}\\
    r^{\pm}_{c} &=& m \pm \sqrt{l^2+m^2} \left(\sqrt{3} \sin \left(\frac{1}{3} \tan^{-1} (\frac{l}{m})\right) \mp \cos \left(\frac{1}{3} \tan^{-1} (\frac{l}{m})\right)\right). \label{eq:solutions2}
\end{eqnarray}
One could see that only the first solution gives a radius larger than that of outer horizon $r^{+}_H$, and at the same time for $l=0$ reduces to the corresponding value for the photon sphere in Schwarzschild black hole, $r_{ps}=3m$ \cite{NPF}. The other two solutions, for all values of $l$ and $m$ are inside the horizon, and are not either physically acceptable, or suitable for our simulation purposes which only cover the exterior of the NUT hole. Also it is noted that both $\pm l$ give the same solution as expected from \eqref{L8}. This is another difference compared to the Kerr case in which the $\pm a$ give two different photon rings as the direct, and retrograde orbits in its equatorial plane \cite{Chandra}.\\
Now that we have the photon ring radius in Schwarzschild-type coordinates, we can transform it to the isotropic coordinates by transforming the equatorial NUT metric to the same coordinates.
The equatorial NUT metric \eqref{n4} can be transformed into the isotropic coordinates by the following transformation,
\begin{equation}
    r = \frac{(2\rho + m)^2 + l^2}{4\rho}
\end{equation}
leading to
\begin{eqnarray}\label{iso}
              && ds^2 =\  f(r(\rho)) \, dt^2 - F(\rho) dl^2_f, \\
            && f(r(\rho)) = \frac{\left(4 \rho ^2-l^2-m^2\right)^2}{l^4+2
   l^2 \left(m^2+4 m \rho +12 \rho
   ^2\right)+(m + 2 \rho )^4} ,\\
            && F(\rho) = \frac{1}{\rho^2} \Bigg[ l^2 + \left( \left(\frac{l}{2\rho} \right)^2 + \left(1+\frac{m}{2\rho} \right)^2 \right)^2 \Bigg]
\end{eqnarray}
where $dl^2_f= d\rho^2 + \rho ^2 d\phi^2$ is the spatial line element in flat spacetime, and the corresponding refractive index for the equatorial NUT will be \cite{NPF},
\begin{equation}\label{RefI}
    n_{\text{\tiny{NUT}}} (\rho) =\ \Bigg[\frac{F(\rho)}{f(r(\rho))}\Bigg]^{\frac{1}{2}} \;  = \frac{1}{4} \frac{l^4+2 l^2
   \left(m^2+4 m \rho +12 \rho
   ^2\right)+(m+2 \rho )^4 }{\rho ^2 \left(4 \rho^2 - l^2 - m^2\right)}
\end{equation}
Locations of the horizon $\rho_H^{+}$,  and the photon ring $\rho_c$ in isotropic coordinates are given respectively by
\begin{equation}\label{rhoH}
    \rho_H^{+} =\ \frac{1}{2} \sqrt{m^2 + l^2}\; .
\end{equation}
and
\begin{eqnarray}\label{rohc}
   \rho_{c} &=&\ \sqrt{l^2+m^2} \cos \left(\frac{1}{3} \tan^{-1} (\frac{l}{m})\right) \\ \nonumber
   &+&\frac{1}{2}
   \sqrt{ \left(\left(l^2+m^2\right) + 2\left(l^2+m^2\right) \cos \left(\frac{2}{3} \tan^{-1} (\frac{l}{m})\right)\right)}.
\end{eqnarray}
The impact parameter of the rays forming the photon ring $b_c$, being a constant of motion, is  given by replacing for $r_{c}$ from \eqref{L9} in \eqref{L7}, which is a  function of $m$ and $l$ as follows,
\begin{equation}
  b_c = \sqrt{\frac{2\left[m+2 \sqrt{l^2+m^2} \cos \left(\frac{1}{3} \tan^{-1} (\frac{l}{m})\right)\right]\left[ {(m+2 \sqrt{l^2+m^2} \cos \left(\frac{1}{3} \tan^{-1} (\frac{l}{m})\right))}^2 + l^2\right]}{2 \sqrt{l^2+m^2} \cos \left(\frac{1}{3} \tan^{-1} (\frac{l}{m})\right)}}.
\end{equation}
Now having both $\rho_{c}$ and $b_c$ we can carry on with the simulation of photon rings. As pointed out previously what we are going to simulate in the next section, is the metamaterial analog of the outer region (i.e $\rho > \rho_H^{+}$) of the NUT equatorial plane. Obviously the assigned index of refractions \eqref{RefI} is also  restricted to the same region.
\section{Ray-tracing simulation of photon rings}\label{SectionIV}
Simulation of light ray trajectories in the metamaterial analog of the NUT equatorial plane with the index of refraction \eqref{RefI} could be carried out in the same way as done for the metamaterial analog of the Schwarzschild black hole in \cite{NPF}. This is so because the equatorial line elements in both cases basically entail the same type of geometry. 
So we use the same relation between the index of refraction and the critical angle the rays should make with the radial direction (i.e angle of the cone of avoidance) to  form the photon ring starting from any given isotropic radial coordinate namely,
\begin{equation}\label{angle}
\sin \Theta_{cr} = \frac{b_{c}}{\rho n(\rho)} \, .
\end{equation}
We use the above equation to simulate light ray trajectories forming the photon rings in the metamaterial analog of the equatorial NUT plane. Details of the simulations are given in appendix A.
Numerical values of mass and NUT factor used in our simulations, leading to different photon rings in  both equatorial NUT and pure NUT ($m=0$) holes,  are presented in table ~\ref{tab:table1}. This table also includes the values for the  photon ring positions obtained both from its exact value \eqref{rohc}, and that obtained from the simulation \footnote{The scales of $m$ and $l$ determine the overall scale of the system and distances included, which could be in centimeters, millimeters, or smaller, depending on the values of the parameters $\{m, l\}$.}. Table~\ref{tab:table1} reveals that the simulation results closely match the exact values, with an agreement at the order of approximately $\sim 10^{-6}$. Achieving this level of precision for the location of photon rings required the light rays to complete at least five orbits around the hole. To attain even greater precision, we must fine-tune the impact parameter $b_{c}$, allowing the rays to complete more than five orbits.
In the subsequent discussions, we will delve into the simulation results for both the NUT and pure NUT spacetimes.
\begingroup
\squeezetable
\begin{table}
\begin{tabular}{|>{\centering\arraybackslash}p{2cm} >{\centering\arraybackslash}p{2cm} >{\centering\arraybackslash}p{2cm}>{\centering\arraybackslash}p{3cm} >{\centering\arraybackslash}p{3cm}|}
\hline
Parameters & $ \rho_{H}$ & $b_{c}$ & $ \rho_{c}^{th}$ & $\rho_{c}^{sim}$\\
\hline
\hline
$m=2,l=0.5$ & 1.0307764 & 10.5699133 & 3.8321034 & 3.8321036 \\
\hline
$m=2,l=2$ & 1.4142136 & 12.7725659 & 5.0695926 & 5.0695928 \\
\hline
$m=2,l=6$ & 3.1622777 & 22.7225471 & 10.6275704 & 10.6275707\\
\hline
$m=0,l=2$ &  1 & 5.6568542 & 3.1462644 & 3.1462645\\
\hline
$m=0,l=6$ &  3 & 16.9705627 & 9.4387931 & 9.4387934\\
\hline
\end{tabular}
\caption[Caption]{The photon ring locations are determined both through simulations and exact calculations for different values of mass and NUT parameters. These photon rings are formed by a light beam originating from an initial distance of $\rho_0 = 36$, while $\rho_c^{th}$ and $\rho_c^{sim}$ are the location of the photon ring from theoretical equations, and the location of the photon ring from simulations, respectively.}\label{tab:table1}
\end{table}
\endgroup
\begin{figure}[h!]
	\centering
	\begin{subfigure}[b]{0.45\textwidth}
		\includegraphics[width=0.85\textwidth]{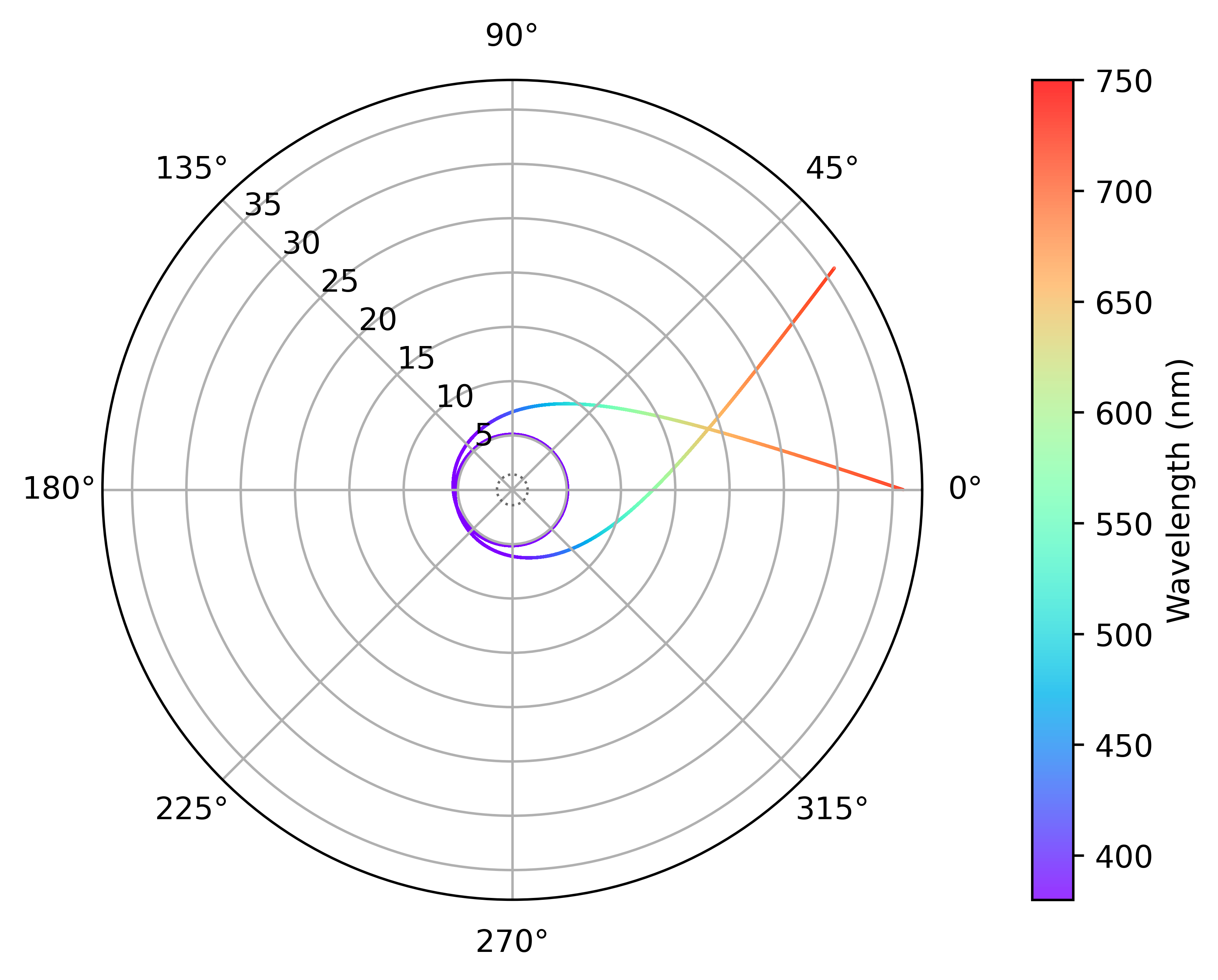}
		\caption{$m=2, l=2$, $\mathcal{W} \approx 2$, $\rho^+_H \approx 1.41$, $\rho_c \approx 5.09$}
		\label{fig:M2L2}
	\end{subfigure}
	\begin{subfigure}[b]{0.45\textwidth}
		\includegraphics[width=0.85\textwidth]{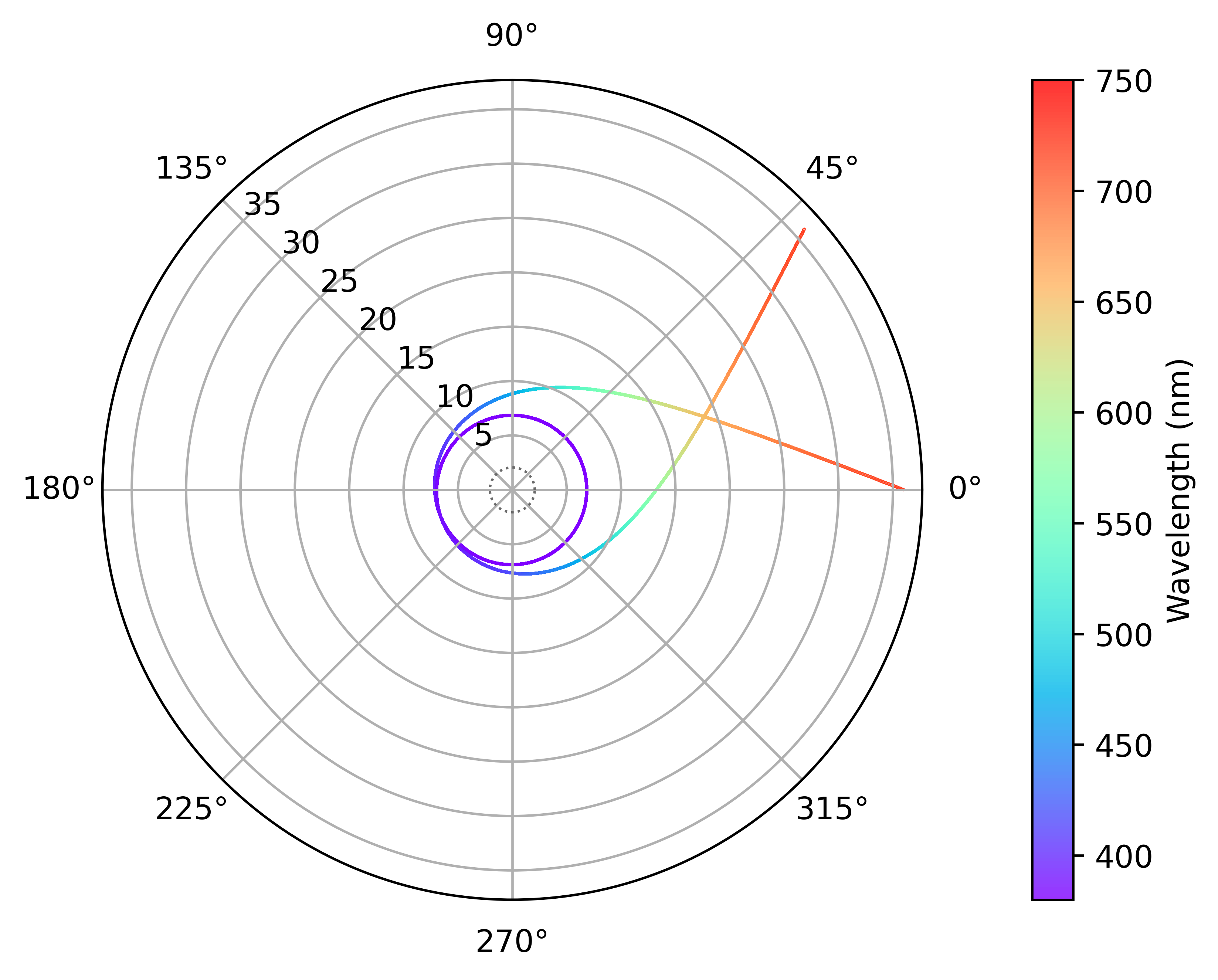}
		\caption{$m=1,l=4$, $\mathcal{W} \approx 2$, $\rho^+_H \approx 2.06$, $\rho_c \approx 6.84$}
		\label{fig:M1L4}
	\end{subfigure}
	\begin{subfigure}[b]{0.45\textwidth}
		\includegraphics[width=0.85\textwidth]{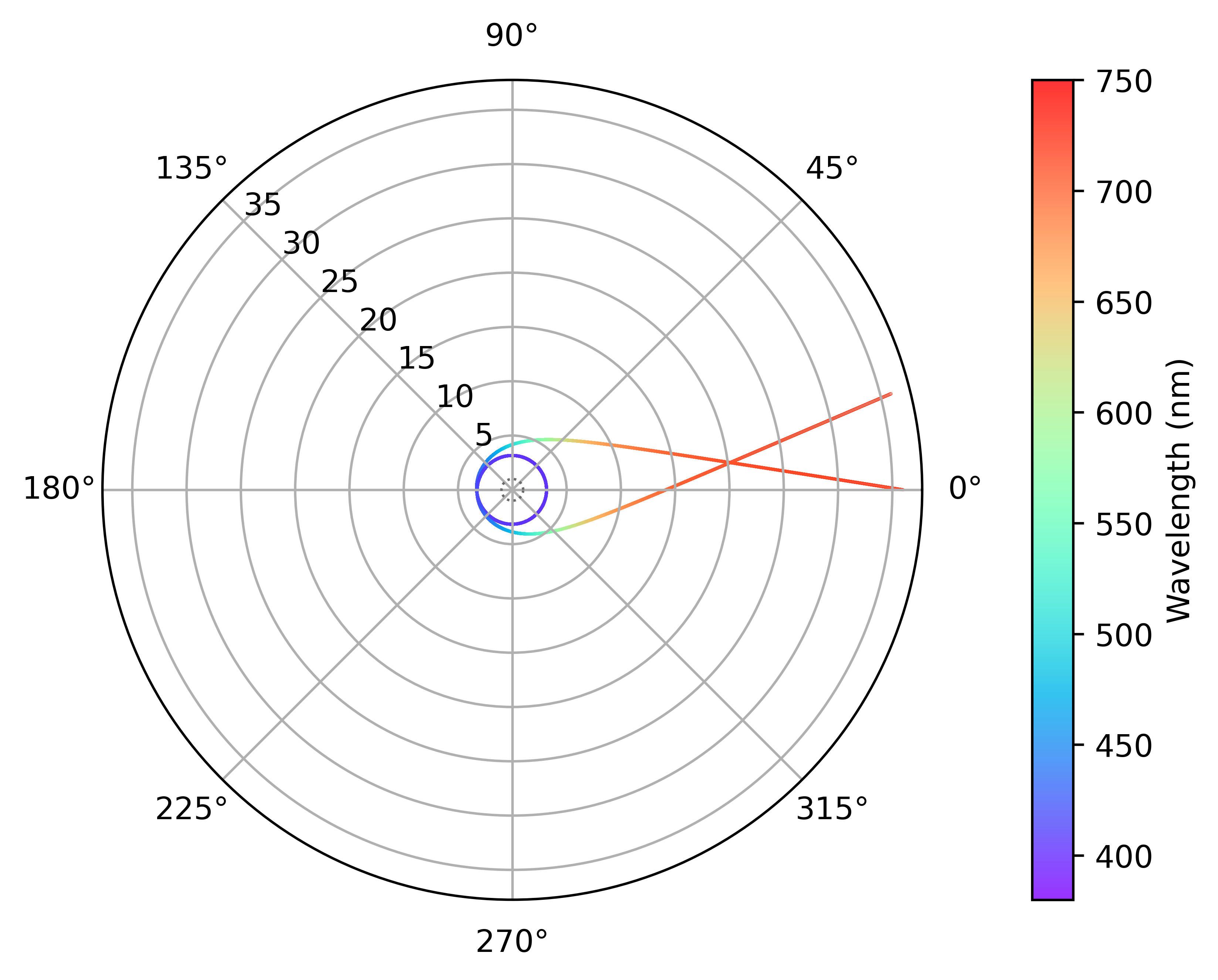}
		\caption{$m=0, l=2$, $\mathcal{W} \approx 2$, $\rho^+_H = 1$, $\rho_c \approx 3.15$}
		\label{fig:M0L2}
	\end{subfigure}
	\begin{subfigure}[b]{0.45\textwidth}
		\includegraphics[width=0.85\textwidth]{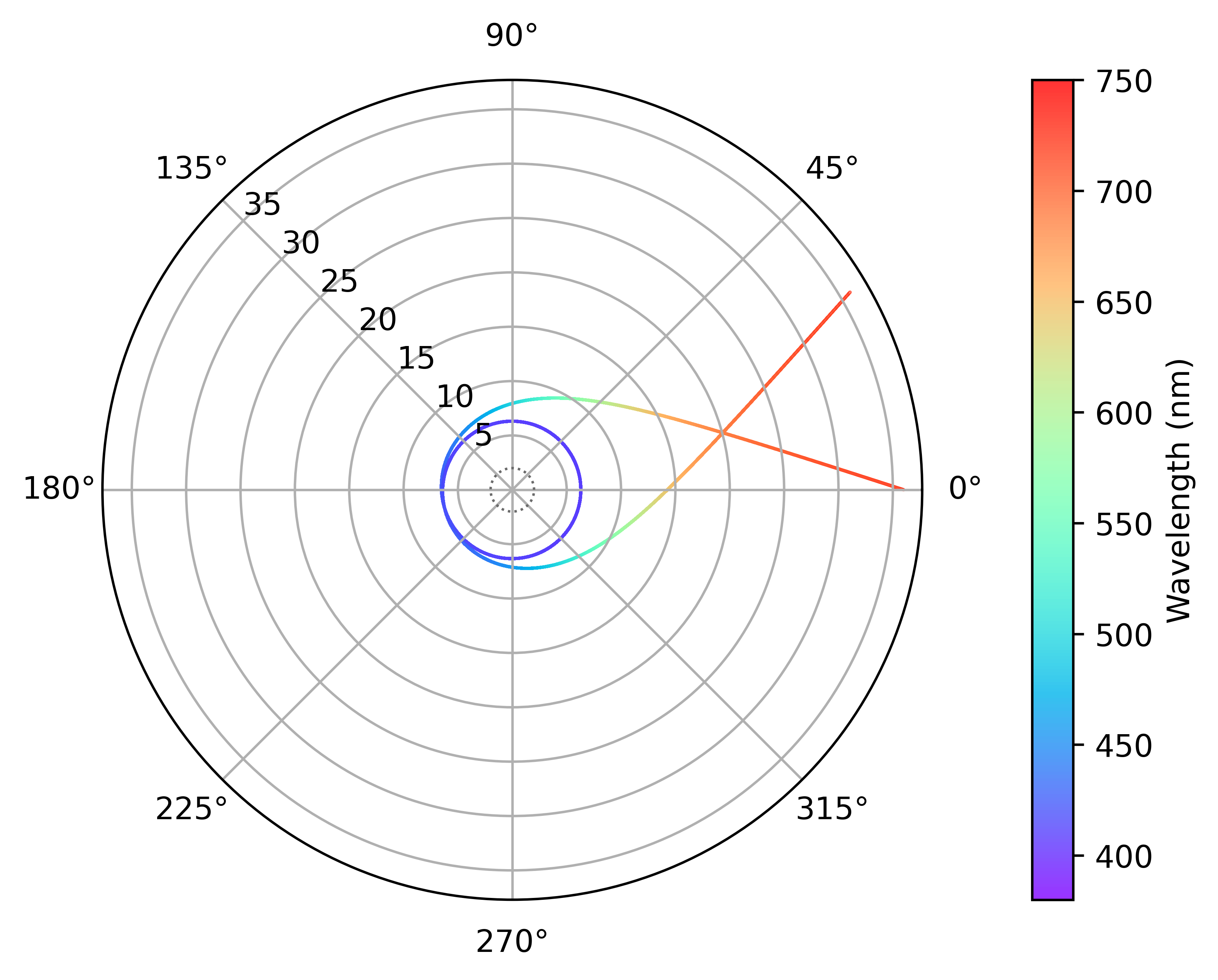}
		\caption{$m=0, l=4$, $\mathcal{W} \approx 2$, $\rho^+_H = 2$, $\rho_c \approx 6.30$}
		\label{fig:M0L4}
	\end{subfigure}
	\caption{Light ray trajectories in the metamaterial analog of the equatorial NUT (a,b), pure NUT (c,d) and charged NUT (e,f) holes, illustrating the formation of
 photon rings (blue and purple circles), while $\mathcal{W}$ is the winding number.}\label{fig:Simu-PR}
\end{figure}

\subsection{The NUT hole}
The results of simulation for a congruence of light ray trajectories in the metamaterial analog of the equatorial NUT hole ($m,l \neq 0$),
leading to the formation of photon rings, are shown in Figs.~\ref{fig:M2L2} and \ref{fig:M1L4} for different values of parameters $m$ and $l$. As is evident, we can change the locations of horizon, and photon rings by changing these parameters. The color palettes in Fig.~\ref{fig:Simu-PR} represent the strength of the refractive index, helping to track its changes as the rays approach the analog photon ring. This visualization also aids in analyzing how the refractive index in the metamaterial domain varies with changes in the free parameters $\{m, l\}$. Consequently, we can easily investigate the different effects that each of these free parameters has on the optical features of the metamaterial medium. The computed wavelength is governed by the relation $\frac{n(\rho_2)}{ n(\rho_1)} = \frac{\lambda(\rho_1)}{\lambda(\rho_2)}$, in the cross section of two consecutive annuli. In the special case of a NUT hole with $m=l=2$, simulation details, and characteristics of the photon ring are tabulated in table  ~\ref{tab:nut}. These include the employed precision of the critical angle $\Theta_{cr}$, the  number of full rotations of rays around the hole (winding number), photon ring position ($\rho_c$), and the difference between photon ring positions  obtained from the theory and the simulation.
\begingroup
\squeezetable
\begin{table}
\begin{tabular}{|>{\centering\arraybackslash}p{1cm} >{\centering\arraybackslash}p{2cm} >{\centering\arraybackslash}p{1cm}>{\centering\arraybackslash}p{2cm}|}
 \hline
$\Delta \Theta_{cr}$ & $\mathcal{W}$ & $\rho_c$  &  $\rho_c^{th}-\rho_c^{sim} $
\\
\hline
\hline
 $10^{-4}$ &  1.5 & 5.2 &  $\sim 10^{-1}$ \\
 \hline
 $10^{-5}$ &  2.2 & 5.08 &  $\sim 10^{-2}$ \\
 \hline
  $10^{-7}$ &  2.9 & 5.071 &  $\sim 10^{-3}$ \\
  \hline
 $10^{-10}$ &  3.7 & 5.0697 & $\sim 10^{-4}$ \\
 \hline
 $10^{-12}$ &  4.7 & 5.06959 &  $\sim 10^{-5}$ \\
 \hline
\end{tabular}
	\caption{The photon ring characteristics of a metamaterial analog of a NUT hole with $m=l=2$ (depicted in Fig.~\ref{fig:M2L2}). Where $\Delta \Theta_{cr}$ and $\mathcal{W}$ represent the precision of the critical angle and the winding number, respectively.}
	\label{tab:nut}
\end{table}
\endgroup
\subsection{The pure NUT hole}
\begin{figure}[b]
		\includegraphics[width=0.8\textwidth]{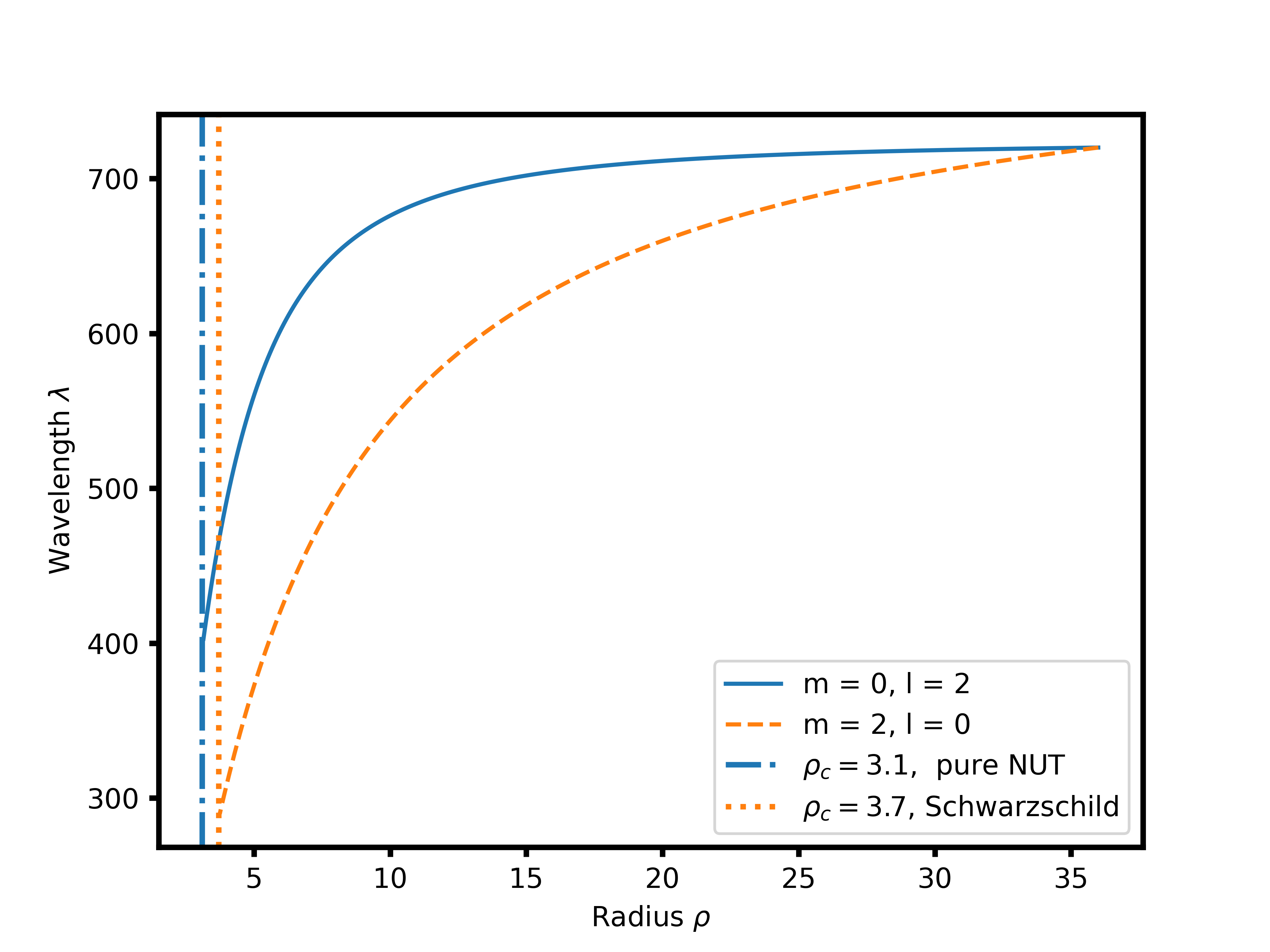}
		\caption{Wavelength profiles for Schwarzschild and pure NUT cases.}
		\label{fig:wavelength}
\end{figure}

In this subsection, our attention is directed towards the pure NUT case, allowing us to explore the exclusive impact of the NUT factor on the optical properties of the corresponding metamaterial analogs. The simulation results for ray trajectories in a metamaterial analog of a pure NUT hole, characterized by NUT factors $l=2$ and $l=4$, while $m=0$ in refractive index \ref{RefI}, $n_{\text{\tiny{Pure-NUT}}} (\rho)$, are depicted in Figs.~\ref{fig:M0L2} and \ref{fig:M0L4}. Comparing these figures  with those for a NUT hole with a non-vanishing mass, Figs.~\ref{fig:M2L2} and \ref{fig:M1L4}, could give us a better insight into the optical effects of the NUT factor $l$ .\\
For example comparing Figs.~\ref{fig:M0L4} with  ~\ref{fig:M1L4}, we observe that both cases have almost the same horizons with differing photon ring radii of about $\sim 2$, while the corresponding refractive indices at the same radii differ notably. Thus, as expected, we find that the presence of a mass parameter increases the refractive index at a given radius, and consequently leads to a decrease in the light wavelength. In contrast, the NUT factor $l$ primarily affects the horizon, and the photon ring positions, with minimal impact on the refractive index and light wavelengths.\\
To better understand how the wavelength changes over the isotropic radius, and investigate the different behaviors promised by each parameter $m$ and $l$, we provide Fig.~ \ref{fig:wavelength}. This figure compares two situations: case I) $\{m = 0, l = 2\}$ (solid blue curve) and case II) $\{l = 0 , m = 2\}$ (dashed orange curve). For these two cases, the locations of the photon rings are indicated by vertical lines (blue dashed-dot for case I and orange dots for case II). This figure also clarifies our point from a wavelength perspective: these two cases have very close locations for the photon ring, while the changes in wavelength and the minimum wavelength in each case are significantly different.
Another noteworthy point is that in all our simulations in Fig.~\ref{fig:Simu-PR}, the maximum winding number for the photon rings is 2, while it is possible to achieve higher winding numbers by increasing computational precision and cost, as shown for the special case of a NUT hole with $m=l=2$ in table ~\ref{tab:nut}.\\
One should be careful not to associate the structure of our metamaterial analog of the equatorial NUT hole with the embedding diagram of the equatorial NUT hole. Indeed our metamaterial analog of the equatorial NUT hole is a flat 2-dimensional surface with the index of refraction adapted from the NUT spacetime geometry, and this is  different from those studies in which spatially curved metamaterials are considered \cite{azevedo2021optical, dos2022simple, atanasov2021wormhole}. These studies necessitate the use of structured metamaterials with curved geometries to achieve an effective refractive index capable of facilitating light-trapping orbits. For a matter of comparison, in appendix B we have simulated light ray trajectories on the embedding diagram of the equatorial NUT hole.
\section{Metamaterial analog of a charged NUT hole as an optical Device: Wave optics approach}\label{waveoptics}
\begin{figure}[h!]
	\centering
	\begin{subfigure}[b]{0.50\textwidth}
		\includegraphics[width=0.8\textwidth]{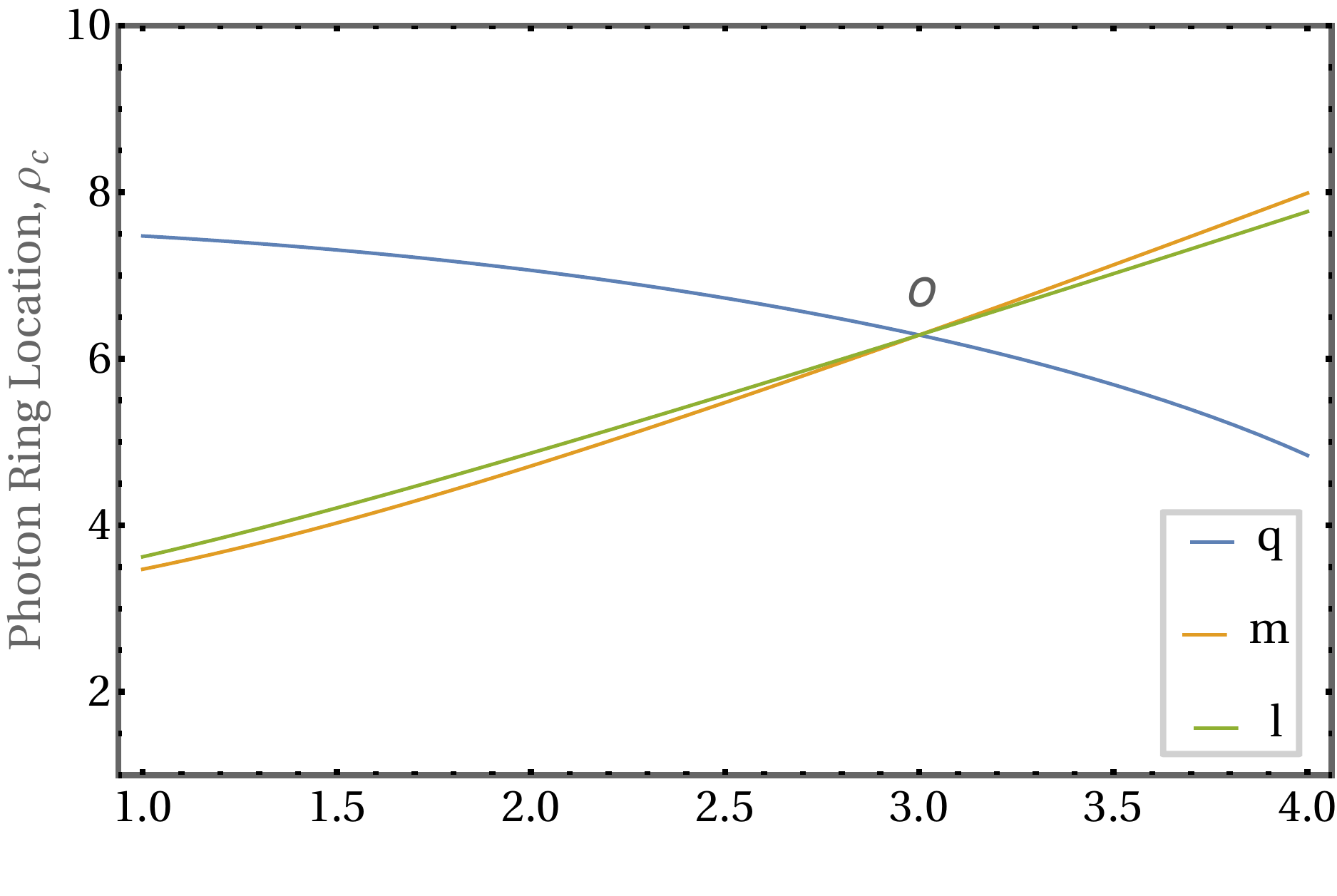}
		\caption{$ O : \{m = q = l = 3\}$}
		\label{fig:PS}
	\end{subfigure}
	\begin{subfigure}[b]{0.50\textwidth}
		\includegraphics[width=0.8\textwidth]{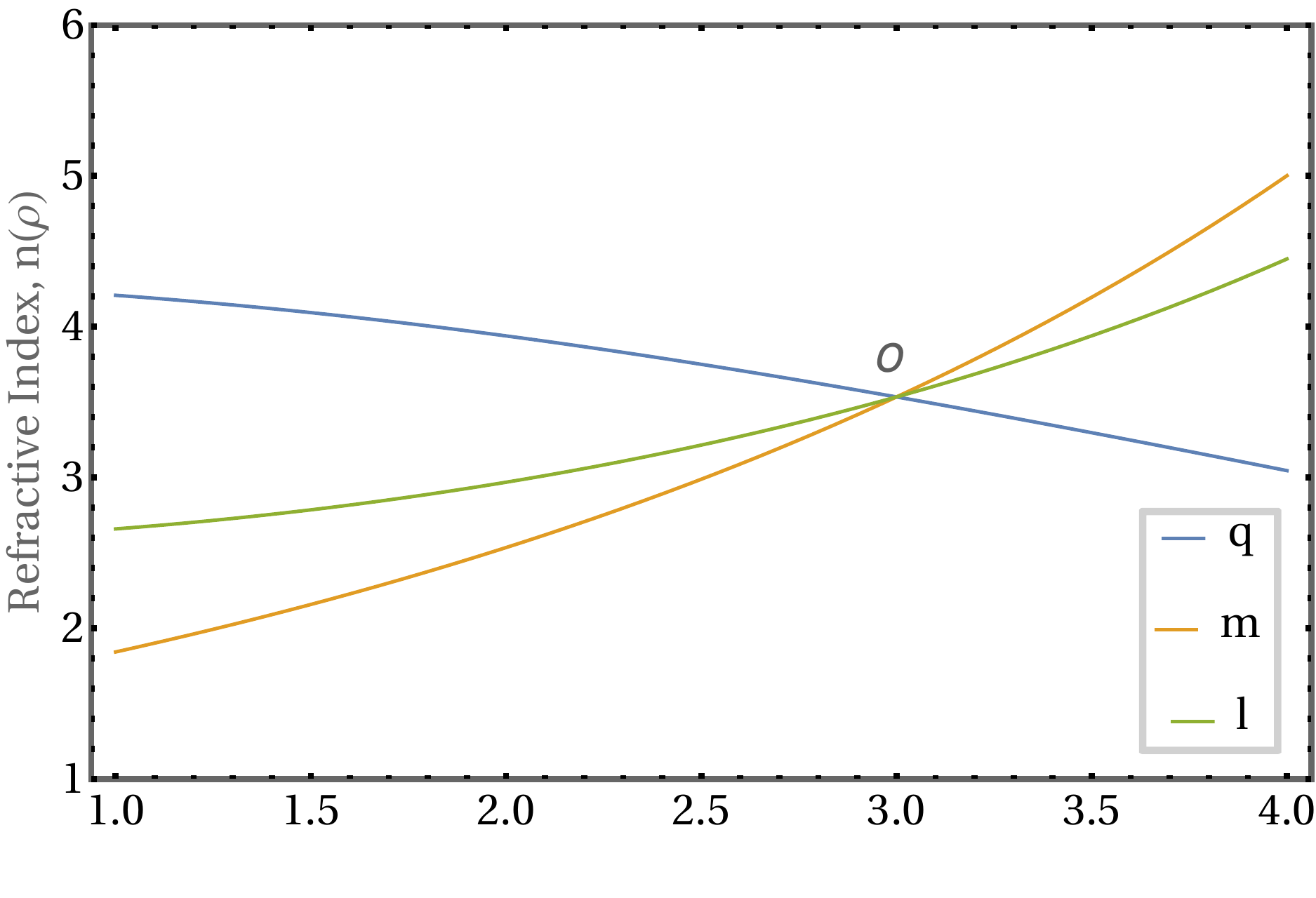}
		\caption{$ O : \{m = q = l = 3\}$}
		\label{fig:n-mql}
	\end{subfigure}
	\caption{Figures (a) and (b) show how $\rho_c$ and $n(\rho)$ change respectively as we move in each direction of the 3D parameter space of the simulation, starting from a fixed point $O : \{m = q = l = 3\}$.}\label{fig:curves}
\end{figure}

\begin{figure}
	\centering
	\begin{subfigure}[b]{0.45\textwidth}
		\includegraphics[width=1\textwidth]{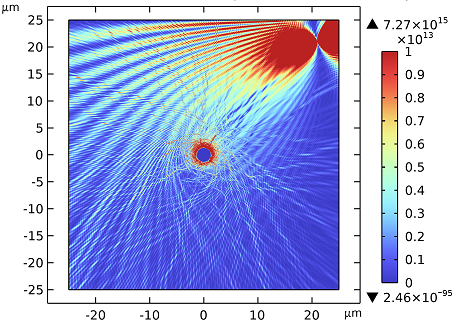}
  		\caption{}
		\label{schwarzschild1}
	\end{subfigure}
	\begin{subfigure}[b]{0.45\textwidth}
		\includegraphics[width=1\textwidth]{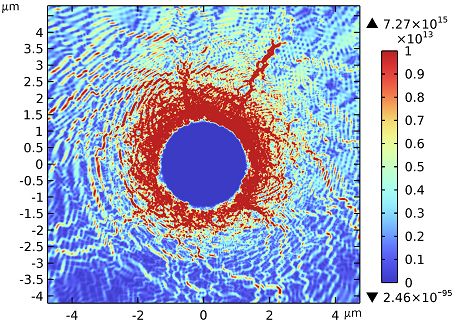}
  		\caption{}
		\label{schwarzschild2}
	\end{subfigure}
	\begin{subfigure}[b]{0.45\textwidth}
		\includegraphics[width=1\textwidth]{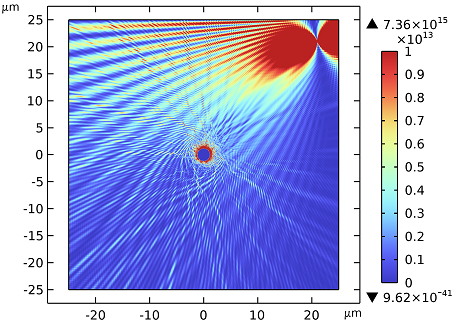}
  		\caption{}
		\label{pureNUT1}
	\end{subfigure}
	\begin{subfigure}[b]{0.45\textwidth}
		\includegraphics[width=1\textwidth]{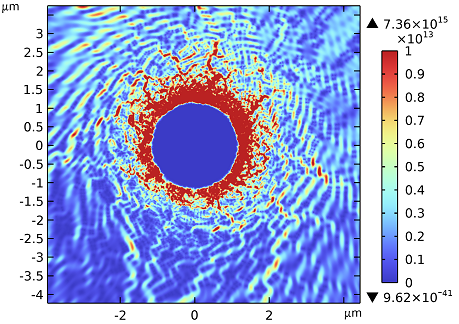}
  		\caption{}
		\label{pureNUT2}
	\end{subfigure}
 	\begin{subfigure}[b]{0.45\textwidth}
		\includegraphics[width=1\textwidth]{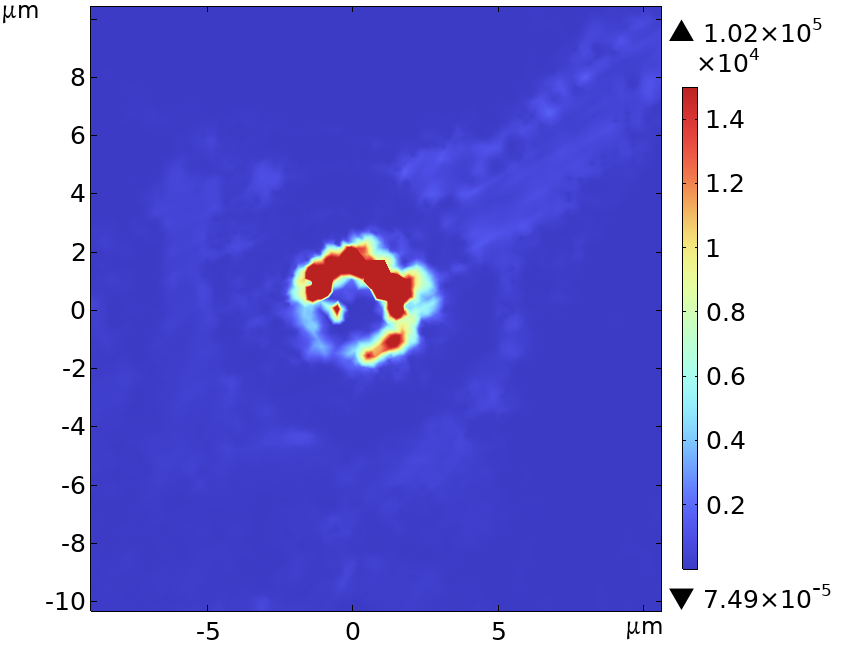}
  		\caption{}
		\label{errorE}
	\end{subfigure}
	\caption{Density plot of the electromagnetic radiation power from an electric dipole oscillating at frequency $f \simeq 6 \times 10^{5}$ GHz in a  metamaterial simulating 1)-Schwarzschild spacetime: (a) and (b), and 2) a pure NUT spacetime: (c) and (d). Plot (e) depicts the density plot of errors in electric field.}\label{fig:WAVE1}
\end{figure}

\begin{figure}
	\centering
	\begin{subfigure}[b]{0.48\textwidth}
		\includegraphics[width=1\textwidth]{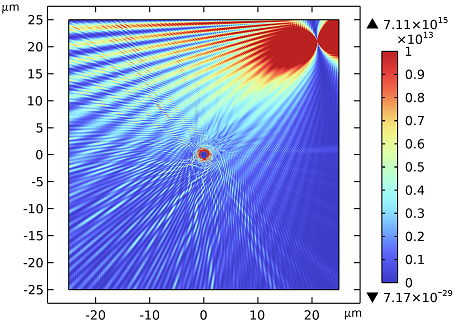}
  \caption{}
  \label{REVERSE1}
	\end{subfigure}
	\begin{subfigure}[b]{0.48\textwidth}
		\includegraphics[width=1\textwidth]{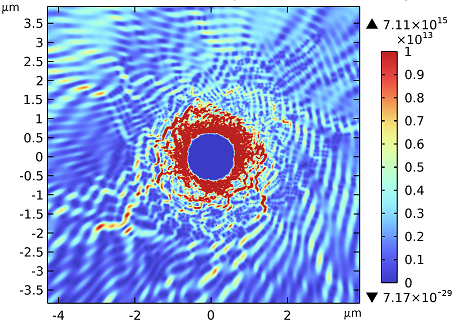}
  \caption{}
  \label{REVERSE2}
	\end{subfigure}
	\caption{Density plot of the electromagnetic radiation power from an oscillating electric dipole at frequency $f = 5.9958$ GHz in a metamaterial mimicking the charged NUT spacetime. Fig. (b) is the same as Fig. (a), but with a higher resolution around the analog horizon.}\label{fig:REVERSE}
\end{figure}

It was noticed in the above simulations that $m$ and $l$ are just two parameters in the metamaterial's isotropic index of refraction Eq. \eqref{RefI}.
Since the  metamaterial analog of a NUT hole  has two different parameters ( $m$ and $l$), one has more control over the  design of the corresponding metamaterial  with the required optical characteristics, as compared to the metamaterial analog of the pure NUT hole which has only one parameter. Utilizing unstable photon rings in metamaterials for the design of optical devices, such as an optical switch, this variety of parameters will allow designers to fine-tune the optical characteristics of the device with enhanced precision.\\
On the other hand, since in our approach, based on the spacetime index of refraction, we can perform an exact ray tracing simulation in the analog metamaterial, by adjusting the winding number of the photon ring, one can significantly increase the sensitivity of the device. This is achieved by increasing the critical angle's precision in the simulation, as detailed in Table ~\ref{tab:nut}.\\
In light of these considerations, this section will focus on the {\it charged} NUT solution, and its equatorial metamaterial analog associated with the spacetime's index of refraction which contains three parameters.
The charged NUT spacetime is an exact solution of the Einstein-Maxwell equations, which is obviously not a vacuum solution. It is noticed that its exotic features, as in the case of NUT solution, all are rooted in its  NUT factor.\\
In Schwarzschild-like coordinates its metric is given by the following line element \cite{Exact},
\begin{equation}\label{cnut}
ds^2 = f(r) (dt - 2l \cos\theta d\phi)^2 - \frac{dr^2}{f(r)} - (r^2 + l^2) d\Omega^2 \, ,
\end{equation}
with
\begin{equation}\label{f1}
f(r) = \frac{r^2 - 2mr - l^2 + q^2 }{r^2 + l^2}\, ,
\end{equation}
in which $m$, $l$ and $q$ are the mass, NUT parameter, and the electric charge respectively. Applying the same procedure used in section \ref{sectionIII}, one can show that the equivalent index of refraction for the equatorial charged NUT is given by,
\begin{equation}\label{RefII}
    n_{\text{\tiny{cNUT}}} (\rho) = \frac{1}{4} \frac{l^4+2 l^2
   \left(m^2+4 m \rho -q^2+12 \rho
   ^2\right)+(m-q+2 \rho )^2 (m+q+2 \rho
   )^2}{\rho ^2 \left(q^2+4 \rho^2 - l^2 - m^2\right)} \, .
\end{equation}
\begin{figure}[h!]
	\centering
\begin{subfigure}[b]{0.45\textwidth}
		\includegraphics[width=0.85\textwidth]{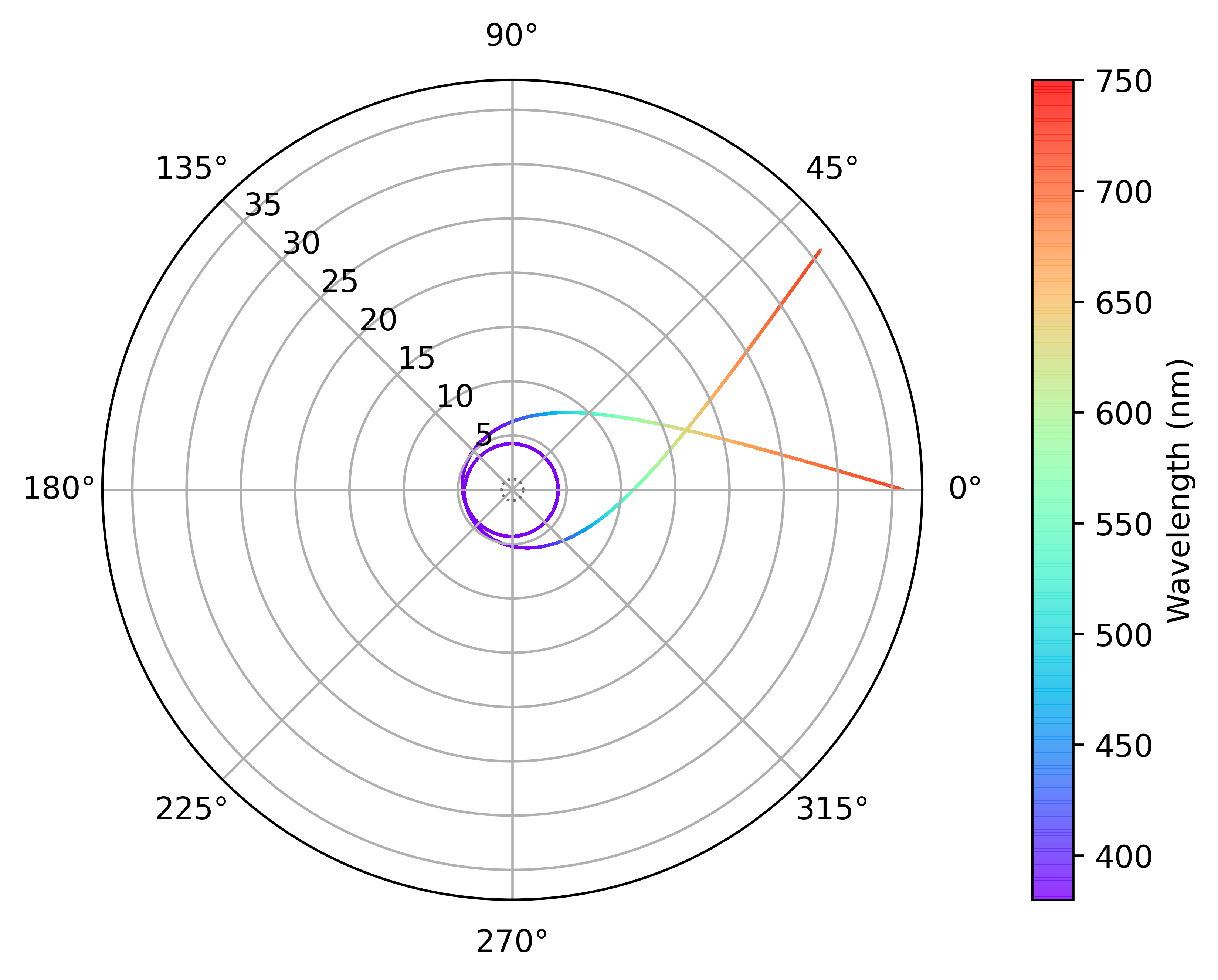}
		\caption{$m=q=l=2$, $\mathcal{W} \approx 2$, $\rho^+_H = 1$, $\rho_c \approx 4.21$}
		\label{fig:M2Q2L2}
	\end{subfigure}
	\begin{subfigure}[b]{0.45\textwidth}
		\includegraphics[width=0.85\textwidth]{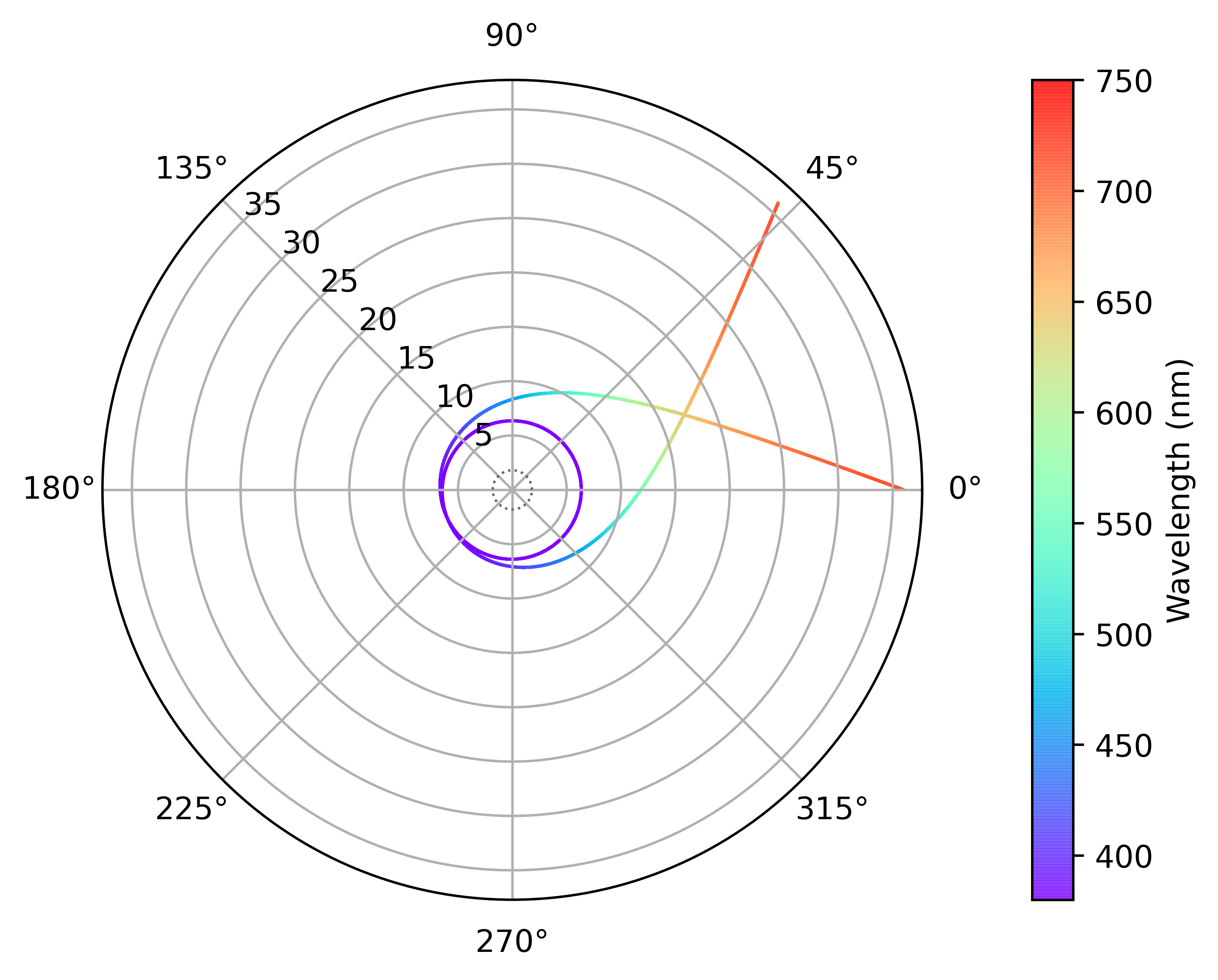}
		\caption{$m=1,q=2, l=4$, $\mathcal{W} \approx 2$, $\rho^+_H \approx 1.80$, $\rho_c \approx 6.33$}
		\label{fig:M1Q2L4}
	\end{subfigure}
\caption{Light ray trajectories in the metamaterial analog of the equatorial charged NUT hole for two different sets of values for ${m,l \;{\rm and}\; q}$.}\label{fig:Simuq-PR}
\end{figure}
Its detailed derivation is outlined in the appendix C. Employing the above index of refraction, the ray-tracing simulation in the metamaterial analog of the equatorial plane of this three-parameter spacetime, for two different sets of values for $m, l \;{\rm and}\; q$ are given in  Fig.~\ref{fig:Simuq-PR}. Comparison of $\rho^+_H$ and $\rho_c$ in these figures with Figs.~\ref{fig:M2L2} and ~\ref{fig:M1L4} which share the same values of $m$ and $l$  show the effect of the charge parameter very clearly.  \\
This three-parameter refractive index enhances the design flexibility of the corresponding metamaterial analog, offering more options for the placement of the analogs of the horizon and the photon ring within the optical device.
This fact is more obvious in Figures~\ref{fig:PS} and \ref{fig:n-mql} which show how the photon ring location and refractive index respond to changes in each of the three parameters \(\{m, q, l\}\) while the other two are kept constant. \(O\) represents a point in the three-dimensional parameter space where \(\{m = q = l = 3\}\). Each curve—blue (only \(q\) varies), green (only \(l\) varies), and orange (only \(m\) varies)—demonstrates how \(\rho_c\) and \(n(\rho)\) change when we move in each direction, while the other two directions remain fixed. Figure~\ref{fig:PS} shows that \(m\) and \(l\) have very similar effects on the location of photon rings when varied. In contrast, Figure~\ref{fig:n-mql} reveals that these two parameters, \(m\) and \(l\), have significantly different effects on the refractive index and, consequently, on the wavelength. Specifically, \(l\) has a lesser effect on \(n(\rho)\) compared to parameter \(m\).\\
In the previous sections, we have used geometric optics, and ray-tracing simulation to investigate the optical properties of the  metamaterial analog of the equatorial NUT spacetime. In this section, we employ wave optics to study the wave behavior of a metamaterial designed with the above index of refraction as a simple optical device.\\
To this end we look for a numerical solution of the Maxwell equations for wave propagation in an inhomogeneous dielectric medium. For our purposes, we assume that the matter is non-magnetic with the index of refraction  $n(\rho) = c \sqrt{\mu_0\, \epsilon(\rho)}$. By combining two of the Maxwell equations, we arrive at the following equation,
\begin{eqnarray}\label{MaxwellE}
    \nabla \times \nabla \times \boldsymbol{E} + \frac{n^2(\rho)}{c^2} \frac{\partial^2 \boldsymbol{E}}{\partial t^2} = 0 \, .
\end{eqnarray}
To replicate the optical behavior of the spacetime of a charged NUT hole in a metamaterial device with the refractive index \eqref{RefII}, we substitute $n_{\text{\tiny{cNUT}}}(\rho)$ for $n(\rho)$.
Fig. \ref{fig:WAVE1} displays the results of numerical simulations for an electric dipole radiating in a medium with refractive index \eqref{RefII} (details of the simulation are given in appendix D). These include two different cases with two different sets of parameters $m, l, q$ : I) $m= 2 \times 10^{-6}, l=q=0$ (the Schwarzschild case, Figs.~\ref{schwarzschild1}-\ref{schwarzschild2}), and II)  $l= 2 \times 10^{-6}, m=q=0$ (pure NUT case, Figs.~\ref{pureNUT1}-\ref{pureNUT2}). Figs. \ref{schwarzschild2} and \ref{pureNUT2}) are the same as  Figs. \ref{schwarzschild1} and \ref{pureNUT1}) respectively, but with higher resolution around their corresponding analog horizons. The event horizons form at $ M/2$, $ l/2$, and the photon rings form at $M/2(2+\sqrt{3})$, $l/2 (\sqrt{2} + \sqrt{3})$ for Schwarzschild and pure NUT cases, respectively. As radiation traverses the metamaterial medium, it curves inward towards the central region, mimicking the behavior of light passing by a massive object, and close to the photon ring position, revolves around the core in nearly-circular orbits.\\
In a few studies, including \cite{Nariman}-\cite{Genov}, Gaussian beams are utilized to explore the wave properties of analog materials. In these studies  $r = 0$ is taken as the singularity point for the proposed refractive index. This obviously does not work for black hole solutions of Einstein field equations. Conversely, Chen et al. \cite{Chen} apply Gaussian beams and truncate their numerical analysis at $r = a$, aiming to more accurately represent a black hole’s event horizon. The isotropic analysis presented in \cite{Fer} employs the exact refractive index of the Schwarzschild metric, revealing wave properties akin to those we have depicted in Figs. \ref{schwarzschild1} and \ref{schwarzschild2}, where even subtle ripples are discernible. Figs. \ref{fig:WAVE1} also illustrate a concentration of electromagnetic radiation around the event horizon, confirming the anticipated analog effect of the spacetime curvature  directing radiation inward towards the center. Despite the fundamental distinctions between wave optics and geometric optics, the findings presented in this section concur with the outcomes detailed in section \ref{SectionIV}.\\
\begin{figure}
\centering
\includegraphics[width=0.8\textwidth]{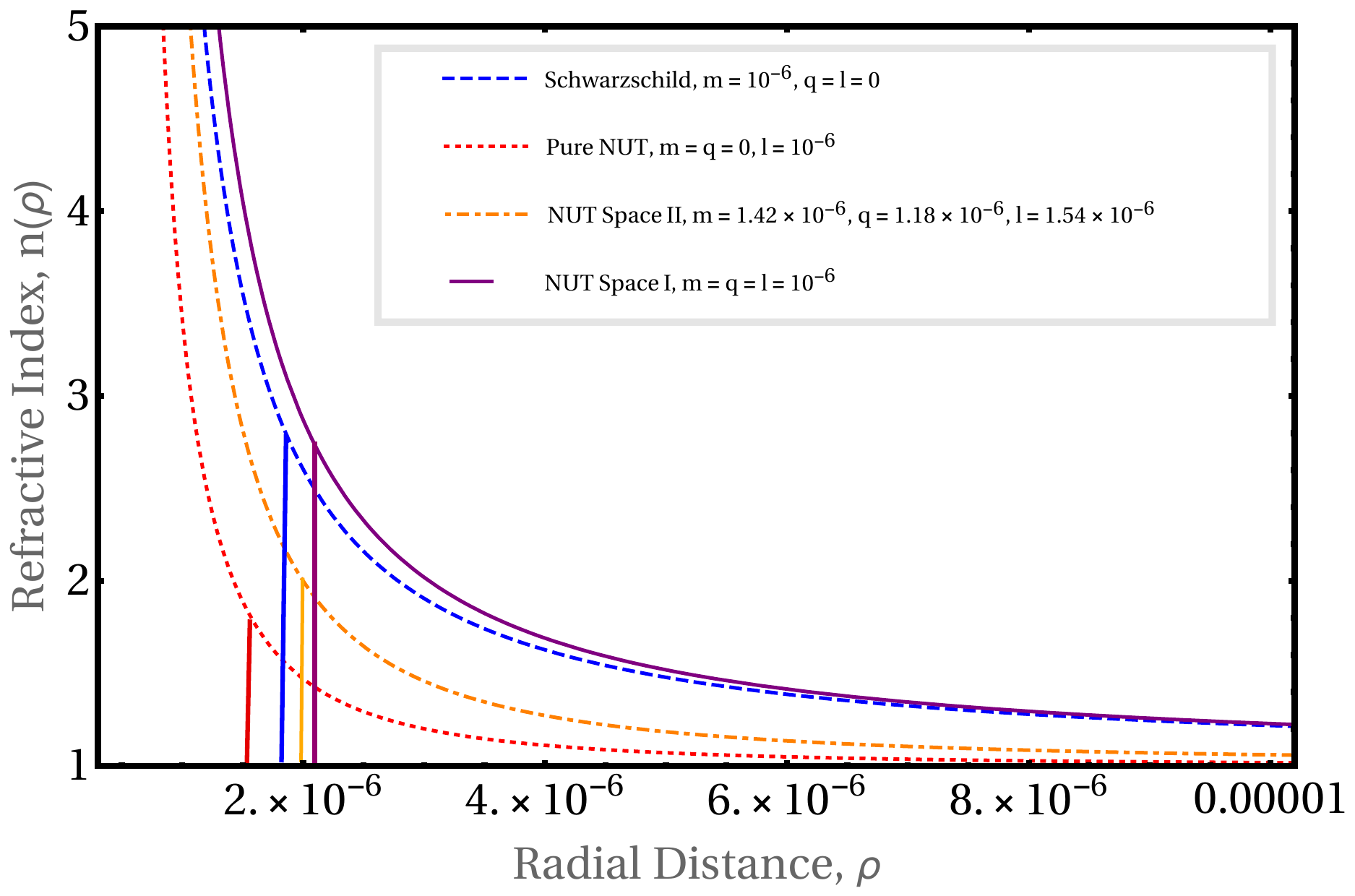}
\caption{Plots of function $n(\rho)$ for different members of the NUT family spacetimes, with vertical lines marking the positions of the photon rings, while all the cases are having the same location for the horizon $\rho_H^+ = 0.5 \times 10^{-6}$.}\label{fig:n-rho}
\end{figure}
As pointed out previously, the interesting optical behavior of the  metamaterial analog of a charged NUT spacetime, with its 3-parameter index of refraction, can be employed in designing novel optical devices.
Fig.~\ref{fig:n-rho} clearly demonstrates that by manipulating the three parameters $\{m, l, q\}$, we gain a greater control over the profile of $n({\rho})$, allowing us to tailor it to specific experimental requirements, including the design of a required optical devices.\\
Alternatively one can obtain the set of the 3 parameters $\{m, l, q\}$ required  for a desired  metamaterial device with a given set of locations for its analog horizon and photon ring, and the value of the refractive index at a given position. For example by locating the horizon and photon ring at $\rho_H = 0.5 \times 10^{-6}$, and $\rho_c = 2 \times 10^{-6}$, respectively, and setting the refractive index value $n(\rho_c) = 2$, a metamaterial device can be crafted with the parameters $m = 1.42 \times 10^{-7},\; l=1.54 \times 10^{-6}$ and, $ q = 1.18 \times 10^{-6}$ \footnote{For this specific set of parameters, the polynomial discriminant $\Delta$ is negative. As expected, among the three potential solutions of the photon ring equation only $r_c$ ($\rho_c$) lies outside the horizon.}. Figure \ref{fig:REVERSE} demonstrates this metamaterial’s optical response to the radiation from an electric dipole. This plot shows a good agreement on the formation of the analogs of the horizon, and the photon ring between the wave optics approach and the ray-tracing simulations carried out in previous sections. It can be seen that as we get closer to the horizon, refractive index (Orange dot-dashed line in Fig.~\ref{fig:n-rho}) increases and works as a barrier for the radiation.
To contextualize our findings in relation to prior research, it is essential to acknowledge that, in order to address the numerical complexities arising from the rapid increase in the profile $n(\rho)$, previous studies have proposed the use of an absorbing inner medium within the core. Specifically, references \cite{Nariman, Chen} advocate a cutoff radius slightly smaller than the horizon's radius, coupled with the assumption of an imaginary absorption medium. This configuration effectively absorbs all incoming waves, thereby emulating the physical characteristics of a black hole's inner horizon. Consequently, the wave concentration observed near the horizon in our simulations does not occur in their configurations. \\
Careful consideration is required when selecting the parameters $\{m, l, q\}$ for a charged NUT-inspired metamaterial device, as the behavior of $n(\rho)$  is critical in our simulations. Theoretically, each parameter set determines a unique horizon position, denoted by \eqref{rhoH}. As one approaches the horizon, $n(\rho)$  increases until diverges at the horizon.
This phenomenon is illustrated in Fig. \ref{fig:n-rho}, where the behavior of the function $n(\rho)$ is depicted for four distinct cases of the NUT family spacetimes: Schwarzschild ($l=q=0, m = 10^{-6}$), pure NUT ($m=q=0, l = 10^{-6}$), charged NUT I ($m = l = q = 10^{-6}$), and the charged NUT II (${m = 1.42 \times 10^{-7}, l = 1.54 \times 10^{-6}, q = 1.18 \times 10^{-6}}$). The vertical lines in the plot mark the corresponding locations of photon rings at  $\rho^{\tiny Sch}_c \simeq 1.86 \times 10^{-6}$,  $\rho^{\tiny PNUT}_c \simeq 1.57 \times 10^{-6}$, $\rho^{\tiny cNUTI}_c \simeq 2.09 \times 10^{-6}$, and $\rho^{\tiny cNUTII}_c = 2\times 10^{-6}$, respectively. These values are in agreement with the numerical solutions presented in Figs. \ref{fig:WAVE1} and \ref{fig:REVERSE}. These cases all share the same horizon location, $\rho_H^+ = 0.5 \times 10^{-6}$,  and were selected to visually demonstrate the flexibility of the three-parameter refractive index \eqref{RefII}. Although they have the same horizon location, the photon rings and the refractive index profiles differ for each case. It should be emphasized that in order to achieve photon rings in optical devices, careful consideration of the value, and profile of $n(\rho)$ near the photon ring is essential. Indeed, a delicate computational meshing needs to be employed around these critical locations, and obviously leveraging a three-parameter refractive index facilitates this manipulation with a greater ease.
\section{Conclusions}
Over the past few decades, analog spacetimes have attracted significant attention for their potential applications both in laboratory experiments and design of optical devices. This study explores the optical properties of the analog spacetime associated with a gravitomagnetic monopole, a subject that has not been previously considered. We have investigated the null ray trajectories, and in particular the formation of photon rings in the metamaterial analogs of the equatorial NUT, and pure NUT spacetimes.
To simulate light rays we have employed a previously introduced relation for the null rays in spherically symmetric spacetimes in their equatorial plane. In this study we have simulated the light ray trajectories in a 2-dimensional metamaterial which mimics the equatorial NUT, pure NUT and charged NUT spacetimes. Assigning a metamaterial with a refractive index identical to that of an equatorial NUT or pure NUT holes in isotropic coordinates, it was shown that the structure of light ray trajectories in these metamaterials exactly mimics that of the corresponding spacetime. This was explicitly shown for the case of  photon rings in the metamaterial analogs of NUT and pure NUT holes. The physical properties of the NUT charge \( l \) in the metamaterial analogs of NUT spaces are investigated through multiple examples and simulations, and compared with the electric charge \( q \), specially, in simulations in Fig. \ref{fig:Simu-PR} and graphs in Figs.~\ref{fig:curves} and \ref{fig:n-rho}.\\
One important observation from this two-parameter metamaterial analog is that, it enables us to identify the location of photon rings, without considerable change in the refractive index, this observation can be confirmed by the provided plots in Figs.~\ref{fig:n-rho} and \ref{fig:curves}. Clearly, as can be seen from $m = 0$, by changing $l$ from 2 to 4, we would have a bigger photon ring, while we do not observe a significant change in the wavelength of the photons, i.e., the increase in the profile of $n_{\text{\tiny{Pure-NUT}}}(\rho)$ remains minimal. This observation can also be confirmed even for the case \( m \neq 0 \) as shown in Fig.~\ref{fig:curves}, where we explored the parameter space of the analog theory and examined the locations of photon rings and horizons.
\\
Most proposals for metamaterial analogs of  black holes in the literature use an effective index of refraction and dielectric permittivity, which are not based on exact solutions of the Einstein field equations \cite{Chen, Leon2, Nariman,  Genov}. These proposals effectively mimic the light trajectories, but not the actual trajectories in the corresponding black hole geometry. In \cite{Fer}, the authors assign a scalar refractive index to the analog medium of Schwarzschild spacetime but do not utilize it directly. Instead, they work with Maxwell’s equations in curved spacetime and, through the eikonal approximation, derive an optical Hamiltonian for light trajectories. In contrast, our study proposes a metamaterial analog of black holes using only a single scalar refractive index. We perform simulations at both the wave and geometric optics limits using this function alone. In an interesting study, the authors in Ref.~\cite{Ting} attempt to go beyond Schwarzschild spacetime and find a metamaterial analog for the Kerr–Newman black hole, presenting a three-parameter refractive index. Our results can be compared to their work from three perspectives: Physically, we have included a different parameter, resulting in a distinct profile for the refractive index. Analytically, we work in isotropic coordinates, unlike their approach which utilizes Schwarzschild-like coordinates (Schwarzschild coordinates in the Schwarzschild metric or Boyer–Lindquist coordinates in the Kerr–Newman metric) to derive the three-parameter refractive index, producing an anisotropic medium. The advantage of our approach is that in isotropic coordinates, the 3-space is conformal to Euclidean 3-space and hence, angles between vectors and ratios of lengths are the same as in Euclidean 3-space, allowing us to use familiar trigonometric relations. While, in Schwarzschild-like coordinates, the radial coordinate ($\rho$ in Ref.~\cite{Ting}) directly measures the circumference of a circle centered on the mass (divided by 2$\pi$) and contains some information about the curvature of 3-space. This observation can be elaborated with the help of a question. Technically, we can perform infinite radial coordinate transformations and derive the corresponding refractive index for the metamaterial analog. The question is, which of these metamaterials mimics the exact null trajectories of the seed black hole? We believe it is the one derived from isotropic coordinates, while others approximately describe the corresponding null geodesics.\\
From the simulation perspective, we use refractive indices \eqref{RefI} or \eqref{RefII} for ray tracing simulations, demonstrating the existence of photon rings in the metamaterial analogs (up to numerical precision), which is absent in their work at both the wave and geometric optics limits. At the wave optics limit, we directly use the three-parameter refractive index in the Maxwell equations \eqref{MaxwellE} to study their physical implications. Additionally, we investigate the advantage of having a three-parameter refractive index for the metamaterial analog of spacetime. By providing an example, we illustrate at the end of Sec.~\ref{waveoptics} with Fig.~\ref{fig:REVERSE} how this feature can aid in designing optical black holes, offering more control over specifying the location of photon ring and horizon, and adjusting the profile of the refractive index $n(\rho)$ around the location of the photon ring.\\
A discussion on the scalability of simulations and system sizes could be practically beneficial:
The optical black holes discussed in the previous section can be scaled from micrometers to centimeters or larger by adjusting the refractive index parameters. This adjustment simplifies both the construction of the optical black hole and the practical expectations of light propagation in the medium, whether in ray or wave optics. The scalability allows for tailoring the system size to meet specific needs for the wavelength and width of the light source, as well as the refractive index gardient. Scaling from micrometers to millimeters or beyond is feasible. The results from wave optics simulations are particularly useful for photonic metamaterials, which can be structured at the nanometer scale to manipulate light at optical frequencies.
In Refs. \cite{azevedo2021optical} and \cite{dos2022simple}, a device made of a nematic liquid crystal, and hyperbolic metamaterial film on a catenoid was proposed. Their effective optical metric may or may not be a solution of the Einstein field equations. On the other hand, instead of using a 2-dimensional curved surface (such as a catenoid), here we propose design of 2-dimensional flat metamaterials endowed with refractive indices adapted from exact solutions of Einstein field equations, in which case the light rays in the metamaterial exactly mimic those in the corresponding spacetime.
We also showed that by employing the same index of refraction in Maxwell's equation and solving them numerically in the wave optics limit, the results are compatible with those obtained from the ray-tracing simulations. Indeed from the figures \ref{fig:WAVE1} one could identify the approximate positions of the photon rings which are compatible with those obtained from the ray-tracing (geometric optics) simulation (refer to the comparable figures in the ray-tracing simulations). \\
In the design of optical devices that replicate the optical characteristics of curved spacetimes, including black hole spacetimes, it is crucial to precisely control the refractive index profile as well as the positions of possible closed photon orbits (spheres and rings), and horizons due to experimental challenges. This level of control is more attainable with  refractive indices having more than one parameter. Also as pointed out in the introduction, those rays passing through the photon ring could be easily handled, and discharged from the metamaterial structure. For example this could be done through a fiber-optic cable along the direction orthogonal to its plane. This is a much easier process, specially when compared with the metamaterial analog of Schwarzschild spacetime, and rays falling through the photon sphere \cite{NPF}.
In some previous studies \cite{Genov, Fer,dos2022simple} the authors made use of a one-parameter refractive index, while our study demonstrated that the three parameter index of refraction adapted from the charged NUT solution, provides enhanced control over the optical characteristics of its analog metamaterial. This is particularly evident in the precise manipulation of the photon ring position, which could be instrumental in the design of advanced optical devices. \\
As a final comment, it is noted from the literature that there are two distinct motivations for investigating metamaterials with refractive indices adapted from black holes: I) To study the optical properties of black holes, as explored in references \cite{Genov,Chen}, and II) To propose optical devices that exhibit unusual features for light, including light-trapping and slow-light capabilities \cite{azevedo2021optical, schurig2006metamaterial, manjappa2015tailoring}.To better understand the utility of the three-parameter refractive index \eqref{RefII} for proposed optical devices, a discussion on several optical device proposals is useful. In the light-trapping devices proposed in \cite{azevedo2021optical, dos2022simple}, a 3-dimensional optical device is suggested to trap light around the throat of the device. However, by employing the refractive index $n_{\text{\tiny{cNUT}}}(\rho)$, we can achieve light trapping on a 2-dimensional surface, eliminating the need for a 3D design. Additionally, the three parameters allow us to adjust the device so that the throat (horizon in our case) and the trapped region (photon rings in our terminology) can be positioned at different locations as desired by the engineers. Notably, $n_{\text{\tiny{cNUT}}}(\rho)$ can also represent the optical concentrator proposed in \cite{azevedo2018optical}. In addition to having a zero radius for the horizon (as utilized in this proposal), $n_{\text{\tiny{cNUT}}}(\rho)$ can also offer a trapped region for light that can be adjusted freely. Furthermore, in proposals like omnidirectional light absorption \cite{Nariman} or optical black-hole cavity devices \cite{QBa2022}, dielectric permittivity $\bm{\epsilon}(r) = \bm{\epsilon}_0 (R/r)^2$ or refractive index $n(r) = n_0 R/r$ are used to concentrate light. These profiles for the optical properties of the device do not offer photon rings and, due to insufficient free parameters, present limitations in the practical design of the device.
For instance, in \cite{Nariman}, the authors examine a spherically symmetric shell with interior and exterior radii denoted as $R_c$ and $R$, respectively. According to their study, the core radius $R_c$ is not an independent parameter but is instead determined by the relation $R_c = R \sqrt{\bm{\epsilon}_0/\bm{\epsilon}_c}$, where $\bm{\epsilon}_c$ and $\bm{\epsilon}_0$ represent the interior and exterior dielectric permittivity.
In contrast, the refractive index $n_{\text{\tiny{cNUT}}}(\rho)$, in addition to offering a photon ring region, is versatile enough to adjust the optical properties of the device at an arbitrary radius, as demonstrated by an example in Sec.~\ref{waveoptics}.\\
Our simulations of the optical behaviour in the metamaterial analog of various NUT hole spacetime were conducted using indices of refraction represented by $n_{\text{\tiny{NUT}}}(\rho)$,  $n_{\text{\tiny{Pure-NUT}}}(\rho)$ and $n_{\text{\tiny{cNUT}}}(\rho)$. These indices were approximated by a concentric circular mesh (with a constant index of refraction in each annulus), suggesting the feasibility of constructing such gradient-index optical analogs with conventional metamaterials. However, to obtain more accurate results, it is necessary to increase the number of annuli. This will provide a more precise simulation, and one could hope to gain a deeper understanding of the optical behavior in black hole geometries through their laboratory analogs, and on the metamaterial side, to explore potential avenues for developing novel metamaterial-based optical devices.
\section*{Acknowledgments}
The authors would like to thank University of Tehran for supporting this project under the grants provided by the research council. They express their gratitude to the Department of Physics, University of Tehran for granting access to its High-Performance Computing (HPC) system, as well as to the School of Physics at the Institute for Research in Fundamental Sciences (IPM) for providing access to its computational facilities. Additionally, they thank E. Kiani for his assistance with the Mathematica software.
This work is based upon research funded by the Iran National Science Foundation (INSF) under the project No. 4005058.
\appendix
\section{Details of ray-tracing simulation}
To simulate light rays in isotropic media, it has been demonstrated that, due to symmetry, all rays are plane curves that satisfy the relation $n r \sin \theta = C$ \cite{NPF}, where \(C\) is a constant that can be determined from the initial firing position, and direction of the ray. Also \(\theta\) is the angle between the radius vector to a point on the light trajectory and the tangent to it at that point. The trajectory equation is given by \cite{NPF}:
\begin{equation}\label{ray3}
\frac{dr}{d\phi} = r\sqrt{\frac{r^2 n^2}{C^2} - 1}.
\end{equation}
For the simulations, the radial increment \(\Delta r\) at each step can be determined using $\Delta r = \frac{dr}{d\phi} (\Delta \phi_0)$, where \(\Delta \phi_0\) is the fixed-step increment of the azimuthal angle, chosen based on the required precision. We employed concentric annuli of constant scalar index \(n(\rho)\) at each simulation step for meshing, with the distance between these circles decreasing at the specific rate \eqref{ray3} as they approach the center. Typically, we set \(\Delta \phi_0 = 10^{-5}\) for our meshing, resulting in an error of order \( |e| \propto 10^{-5} \) in observables such as the locations of the photon ring and horizon. This allows us to better test the results against exact values derived from precise equations. However, for application and manufacturing purposes, we propose using steps of order \(\Delta \phi_0 \sim 10^{-2}\), which results in an error of order \( |e| \propto 10^{-2} \). This configuration will produce a meshing of concentric annuli with radial differences ranging in $ 10^{-7} < |\Delta r|/\rho^+_H < 1$ (where exterior horizon $\rho^+_H$ represents the size of the system and can be adjusted according to design and manufacturing needs). In the geometric optics limit, it is enough to have a laser light source with wavelength $\lambda \ll |\Delta r|/\rho^+_H$, to sufficiently mimic the trajectories we depicted in the ray tracing simulations, performed using a Python program we developed for this purpose.
\section{Ray-tracing on the embedding diagram of NUT hole}
Here show how one can acquire a better visualization of the photon rings in the equatorial NUT hole by embedding them into the Euclidean 3-space. Also it is a kind of consistency check demonstrating that the photon ring trajectories produced in the body of the paper can indeed exhibit the expected behavior when embedded into Euclidean 3-space.
To this end we employ the embedding of the equatorial plane of NUT spacetime into the 3-dimensional Euclidean space with the following metric in cylindrical coordinates \cite{Sadegh},
\begin{equation}\label{ds}
ds^2 =dZ^2 + dR^2 + R^2 d\phi^2,
\end{equation}
\begin{figure}
	\centering
	\begin{subfigure}[b]{0.75\textwidth}
		\includegraphics[width=1\textwidth]{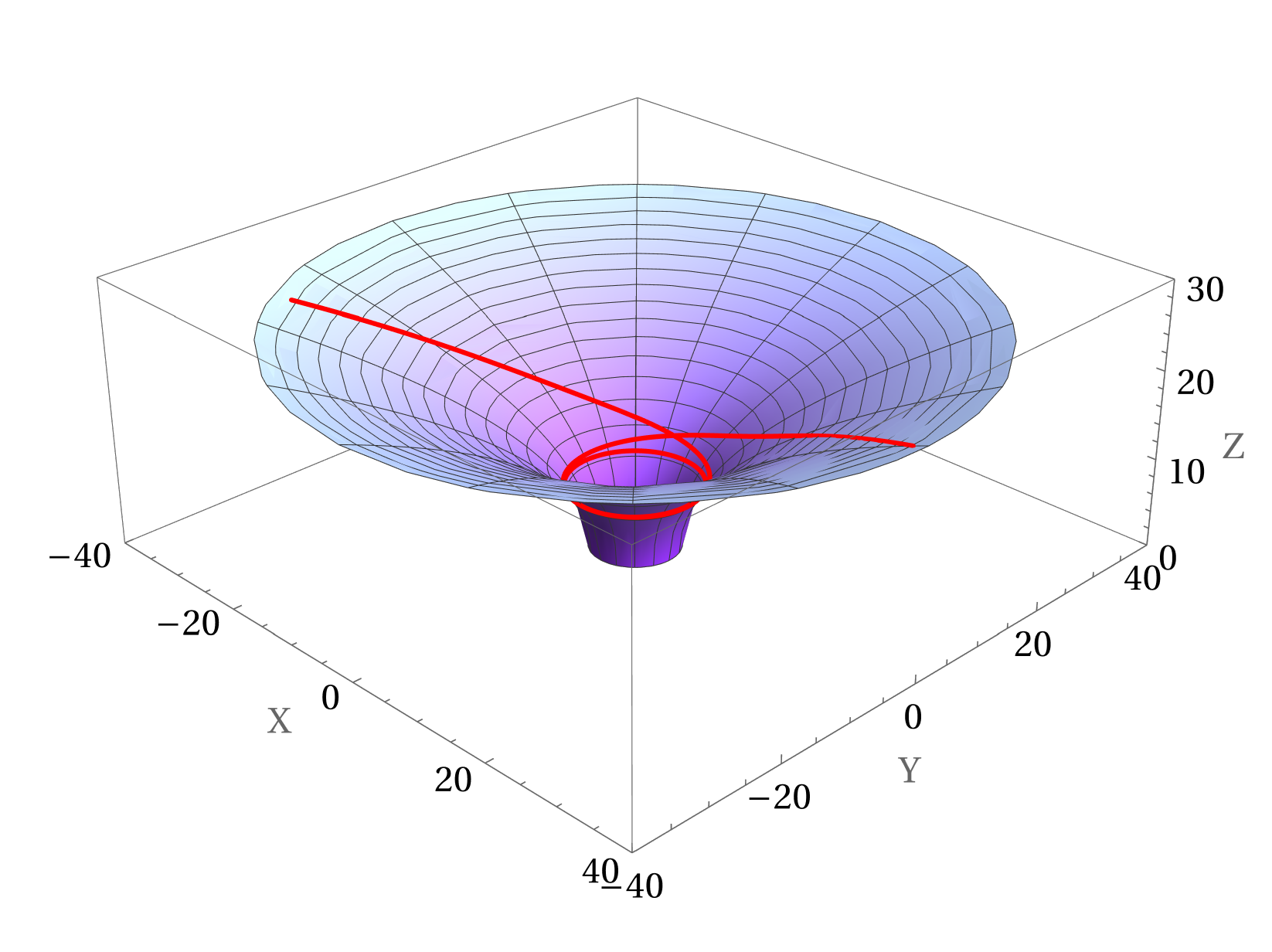}
	\end{subfigure}
	\caption{Simulated light ray trajectory forming a photon ring embedded into the Euclidean 3-space for a NUT hole with $l = m = 2$.}\label{fig:10rays}
\end{figure}
in which, from the angular part of the NUT metric \eqref{n1}, we have $R^2 = r^2 + l^2$ . It is shown that the usual embedding procedure will lead to the following embedding function $Z=Z(R)$ \cite{Sadegh},
\begin{equation}\label{d1}
Z(R) = \int_{R_H}^R \left(\frac{3l^2 \tilde{R}^2 + 2m(\tilde{R}^2-l^2) ^{3/2} -2l^4}{(\tilde{R}^2-l^2)[\tilde{R}^2-2l^2 - 2m(\tilde{R}^2 - l^2)^{1/2}]}\right)^{1/2} d\tilde{R}.
\end{equation}
where the result of integration is valid for  $ R > R_H =({r_H}^2 + l^2)^{1/2}$. Now to find the light ray on the embedding diagram we only need to assign the extra $Z[R(\rho)]$-dimension (obtained from the above integral and, written in terms of the isotropic radius) to any point on the ray  with the isotropic radius $\rho$. The result of ray-tracing simulation on the embedding diagram of a NUT hole with $l = m = 2$ is depicted in Fig. \ref{fig:10rays} for a congruence of 10 rays. These figures show clearly how the light rays wrap around
the throat of the embedding diagram at the radial location of the photon ring.
\section{Charged NUT hole}\label{app:Q-NUT}
In this appendix, we derive the refractive index \eqref{RefII} for a charged NUT spacetime. We begin by substituting \(\eqref{f1}\) into equation \(\eqref{L4}\) to derive the equation for the photon ring,
\begin{equation}\label{A6}
({r_c}^2 + l^2)^2 = ({r_c}^2 -2mr_c - l^2+q^2){b_c}^2,
\end{equation}
as the generalization of Eq.~\eqref{L6} in the presence of charge. By following the same procedure as in the NUT case, we can derive the equation governing the unstable circular photon rings by taking the derivative with respect to $r_c$, which yields Eq.~\eqref{photonring2}.
Thus, the impact parameter $b_c$, will have the same relation as \eqref{L7}.
After substituting $b_c$ back into the equation \eqref{A6}, we derive the following equation governing the position of the photon ring
\begin{equation}\label{A8}
{r_c}^3 - 3 m {r_c}^2 - 3 l^2 r_c + 2 q^2 r_c + m l^2 = 0 .
\end{equation}
If the polynomial discriminant $\Delta = m^2 \left(l^2+m^2-q^2\right)^2-\frac{1}{27} \left(3 l^2+3 m^2-2 q^2\right)^3$ is negative ($\Delta < 0$), all three roots are real and unequal with the following solutions
\begin{eqnarray}
    r_c &=& m+2 \sqrt{l^2+m^2-\frac{2 q^2}{3}} \cos \left(\frac{1}{3} \tan^{-1} (\xi)\right),\label{eq:solutions-RNNUT1}\\
    r^{\pm}_{c} &=& m \pm \sqrt{l^2+m^2-\frac{2 q^2}{3}} \left(\sqrt{3} \sin \left(\frac{1}{3} \tan^{-1} (\xi)\right) \mp \cos \left(\frac{1}{3} \tan^{-1} (\xi)\right)\right), \label{eq:solutions-RNNUT2}
\end{eqnarray}
where 
\begin{equation}
    \xi =\ \frac{\sqrt{\left(3 \left(l^2+m^2\right)-2
   q^2\right)^3-27 m^2
   \left(l^2+m^2-q^2\right)^2}}{3 \sqrt{3} m
   \left(l^2+m^2-q^2\right)} \, .
\end{equation}
These solutions represent a generalization of the solutions given by equations \eqref{eq:solutions1} and \eqref{eq:solutions2}, for the case of charged NUT spacetime. As in the NUT case, we observe that only the first solution yields a radius greater than the horizon $r_H$. Simultaneously, when $l= q =0$, it reduces to the corresponding value for the photon ring in the Schwarzschild black hole, i.e., $r_{ps}=3m$. However, the other two solutions, regardless of the values of $l$, $m$ and $q$, lie inside the horizon and are not considered physically valid.\\
Therefore, the equatorial charged NUT spacetime in isotropic coordinates and its refractive index can be derived by applying the following coordinate transformation
\begin{equation}
    r = \frac{(2\rho + m)^2 + l^2 -q^2}{4\rho},
\end{equation}
leading to
\begin{eqnarray}\label{iso}
              && ds^2 =\  f(r(\rho)) \, dt^2 - F(\rho) dl^2_f \, , \\
            && f(r(\rho)) = \frac{\left(q^2+4 \rho ^2-l^2-m^2\right)^2}{l^4+2
   l^2 \left(m^2+4 m \rho -q^2+12 \rho
   ^2\right)+(m-q+2 \rho )^2 (m+q+2 \rho )^2} \, ,\\
            && F(\rho) = \frac{1}{\rho^2} \Bigg[ l^2 + \left( \left(\frac{l}{2\rho} \right)^2 - \left(\frac{q}{2\rho} \right)^2 + \left(1+\frac{m}{2\rho} \right)^2 \right)^2 \Bigg] \, .
\end{eqnarray}
The refractive index corresponding to the equatorial charged NUT  can be expressed as,
\begin{equation}\label{RefI-RNNUT}
    n_{\text{\tiny{RN-NUT}}} (\rho) =\ \Bigg[\frac{F(\rho)}{f(r(\rho))}\Bigg]^{\frac{1}{2}} \;  = \frac{1}{4} \frac{l^4+2 l^2
   \left(m^2+4 m \rho -q^2+12 \rho
   ^2\right)+(m-q+2 \rho )^2 (m+q+2 \rho
   )^2}{\rho ^2 \left(q^2+4 \rho^2 - l^2 - m^2\right)} \, ,
\end{equation}
Based on the given line element, the positions of the horizon, $\rho_H$, and the photon ring in isotropic coordinates are given by
\begin{equation}\label{rhoH}
    \rho_{H}  =\ \frac{1}{2} \sqrt{m^2 + l^2 -q^2}\; .
\end{equation}
and
\begin{eqnarray}
   \rho_{c} &=&\ \sqrt{l^2+m^2-\frac{2q^2}{3}} \cos \left(\frac{1}{3} \tan^{-1} (\xi)\right) \\ \nonumber
   &+&\frac{1}{2}
   \sqrt{ \left(\left(l^2+m^2\right) - \frac{q^2}{3}+\left(2\left(l^2+m^2\right) -\frac{4q^2}{3} \right) \cos \left(\frac{2}{3} \tan^{-1} (\xi)\right)\right)},
\end{eqnarray}
respectively. The impact parameter $b_c$ for the rays forming the photon rings, is a constant of motion depending on the parameters $m$, $l$ and $q$. It can be obtained by substituting the value of $r_{c}$ from equation \eqref{eq:solutions-RNNUT1} into equation \eqref{L7}.
\section{Details of dipole radiation simulation} The electric dipole radiation from an oscillating dipole moment \( |\textbf{p}| \sim 1.02 \times 10^{6} \rho^+_{H}\) at a wavelength of \( \lambda \sim 0.5 \rho^+_{H} \) inside a metamaterial with a refractive index \eqref{RefII} is simulated using the finite difference frequency domain solver in COMSOL Multiphysics for Maxwell’s equations. The simulation employs a tessellation with a minimum and maximum element sizes of \( \sim 10^{-7} \rho^+_{H} \) and \( \sim 0.01 \rho^+_{H} \) respectively for the 2D simulation domain, incorporating scattering boundary conditions to minimize reflection effects from the domain boundaries. Physical quantities are expressed in terms of the horizon radius ($\rho^+_{H}$) to illustrate that the system size can be adjusted based on the free parameters of the refractive index and the size of the optical black hole.
The L2 norm of the error squared, which provides an overall measure of the error distribution throughout the entire solution domain, is presented as a density plot in Fig.~\ref{errorE}. Overall, the errors are \( |e|/|\bold{E}| \leq 10^{-8} \). Each simulated optical black hole is centered in the middle of the 2D domain. The radiation propagates through the metamaterial medium, not directly towards the optical black hole, to better visualize the behavior of the radiation near the center.

\end{document}